\documentclass[preprint,amsmath,amssymb]{revtex4}

\usepackage[dvips]{graphicx}
\usepackage{graphicx}
\usepackage{amsmath}
\usepackage{amssymb}
\usepackage{amsfonts}
\usepackage{upgreek}
\usepackage{txfonts}
\usepackage{color}
\usepackage{latexsym}
\usepackage{amsmath}
\usepackage{bm}
\usepackage{yfonts}

\usepackage{hyperref}
\usepackage{xcolor}
\definecolor{dark-red}{rgb}{0.4,0.15,0.15}
\definecolor{dark-blue}{rgb}{0.15,0.15,0.4}
\definecolor{medium-blue}{rgb}{0,0,0.5}
\hypersetup{
    colorlinks, linkcolor={dark-blue},
    citecolor={dark-blue}, urlcolor={medium-blue}
}

\newcommand{\beq}{\begin{equation}}
\newcommand{\eeq}{\end{equation}}
\newcommand{\ba}{\begin{array}{ccc}}
\newcommand{\ea}{\end{array}}
\newcommand{\nn}{\nonumber \\}
\newcommand{\bx}{{\bm x}}
\newcommand{\bk}{{\bm k}}
\newcommand{\bp}{{\bm p}}
\newcommand{\bq}{{\bm q}}
\newcommand{\bl}{{\bm l}}
\def\bea{\begin{eqnarray}}
\def\eea{\end{eqnarray}}
\begin{document}

\title{\mbox{Conserved current correlators of conformal field theories in 2+1 dimensions}}

\author{Yejin Huh}
\email{yejinhuh@fas.harvard.edu}
\author{Philipp Strack}
\email{pstrack@physics.harvard.edu}
\author{Subir Sachdev}
\email{sachdev@g.harvard.edu}
\affiliation{Department of Physics, Harvard University, Cambridge MA 02138}

\date{\today}

\begin{abstract}

We compute current correlators of the CP$^{N-1}$ field theory in 2+1 dimensions, both at the critical point
and in the phase with spontaneously broken $S\!U(N)$ symmetry. Universal constants are obtained to next-to-leading
order in the $1/N$ expansion. Implications are noted for quantum critical points of antiferromagnets, and their vicinity.

\end{abstract}

\maketitle

-----------------------------------------------------------------------------------------------------------------\\[-15mm]
\tableofcontents

\section{Introduction}

Conformal field theories (CFTs) appear as the low energy description of numerous critical points or phases of interest in condensed matter
physics. Perhaps the best studied are those of two-dimensional insulating quantum antiferromagnets. In dimerized antiferromagnets, we obtain 
`conventional' critical points between N\'eel and spin-gap phases, which are described by a field theory of the N\'eel order at the O(3) Wilson-Fisher
CFT \cite{wenzel}. In antiferromagnets with half-odd-integer spin per unit cell we have the possibility of `deconfined' quantum critical points, described by a CFT of $S=1/2$ bosonic spinons interacting with an emergent gauge field \cite{rsprb,science04}. 
In the latter systems, it is also possible to have `algebraic spin liquid' critical 
phases, described by CFTs of $S=1/2$ fermonic spinons interacting with emergent gauge fields \cite{rantner,rantner02,cfw}. 

Our interest here will be on the correlators of a variety of conserved currents, $J_\mu$, ($\mu$ is a spacetime index) of such CFTs in 2+1 dimensions.
From the general properties of CFTs, we know that the two-point correlator obeys
\beq
\langle J_\mu(-p) J_\nu(p) \rangle = -C \, |\bp| \left(\delta_{\mu\nu}-\frac{\bp_\mu \bp_\nu}{p^2}\right) \label{eq:JJ}
\eeq
where $C$ is a universal number (after some conventional normalization in the definition of $J_\mu$). 
We will present here computations of $C$ to next-to-leading-order in a $1/N$ expansion, where $N$ is the number of {\em flavors\/} of the matter field. Similar computations have appeared earlier for the O($M$) Wilson-Fisher CFT \cite{cha91,fazio96,petkou96}, and for gauge theories with fermionic matter \cite{cfw,rantner02,mastro13}. Our focus will be on the CP$^{N-1}$ model of complex bosonic spinons $z_a$ ($a = 1\ldots N$) coupled to a U(1) gauge field $A_\mu$, 
which describes deconfined quantum critical points in a variety of antiferromagnets \cite{sandvik,kaul12,anders1,white,ganesh,damle2,block13,kaul13}. 

For the case where $\int d^2 x\, J_0$ is the conserved total spin, $C$ is equal to the dynamical spin conductivity measured at frequencies much
larger than the absolute temperature
$\hbar \omega \gg k_B T$. The zero frequency spin conductivity (related by the Einstein relation to the spin diffusivity)
is a {\em separate\/} universal constant \cite{damle}, which we shall not compute here: its computation requires solution of a quantum Boltzmann equation
in the flavor large $N$ limit \cite{ssqhe}, or holographic methods in a matrix large $N$ limit \cite{myers}. 

Turning to the case of the $CP^{N-1}$ field theory, we will consider correlators of its two distinct conserved currents. The first is the SU($N$) flavor current 
\begin{align}
J^\ell_{\mu}=-i z_a^\ast T^\ell_{ab}\left(D_\mu z_b\right) + i \left(D_\mu z_a\right)^\ast T^\ell_{ab} z_b\;,
\label{eq:flavor_current}
\end{align}
where $D_\mu$ is the co-variant derivative, and $T^\ell$'s are generators of the $S\!U(N)$ group normalized 
so that Tr$(T^\ell T^m)=\delta^{\ell m}$. The normalization convention for SU(2) antiferromagnets differs, and its physical total spin current 
is $J^{\ell} /\sqrt{2}$. For the 2-point correlator we have 
\begin{align}
\langle J^\ell_\mu(-p) J^m_\nu(p) \rangle = -C_J \delta^{\ell m}\, |\bp| \left(\delta_{\mu\nu}-\frac{\bp_\mu \bp_\nu}{p^2}\right),
\end{align}
 for which we find the numerical value (see Section.~\ref{subsec:cj_cpn})
\begin{align}
C^{CP^{N-1}}_J = \frac{1}{16}\left( 1 - \frac{2.74}{N}\right)\;.
\end{align}
in the $1/N$ expansion. The second conserved current of the $CP^{N-1}$ theory is the topological current $B_\mu  = 
\epsilon_{\mu\nu\lambda} \partial_\nu A_\lambda$: this measures the current of the Skyrmion spin textures, which is conserved at
asymptotically low energies near deconfined critical points. For this current, we find in Sec.~\ref{sec:ca} a correlator as in Eq.~(\ref{eq:JJ}), with
\begin{align}
C_A = \frac{16}{N}\left( 1+\frac{0.578}{N}\right)\;.
\label{eq:C_A_intro}
\end{align}

It is interesting to compare these results with those of the O($M$) Wilson-Fisher CFT of an $M$-component, real scalar field $\phi_a$. Now 
the current is $J^\ell_\mu = -i \phi_a t^{\ell}_{ab} \partial_\mu \phi_b $, the 
generators $t^\ell$ are purely imaginary antisymmetric matrices conventionally normalized as Tr$(t^\ell t^m)=2 \delta^{\ell m}$. In this case $J^\ell_\mu$ is the physical total spin current of the antiferromagnet for $M=3$. We recall the result \cite{cha91}
\begin{align}
C^{O(M)}_J=\frac{1}{16}\left(1-\frac{1}{M} \frac{64}{9\pi^2}\right)
\approx \frac{1}{16}\left(1-\frac{1}{M} 0.72\right)\;;
\end{align}
we will reproduce this result by the methods of our paper in Sec.~\ref{subsec:cj_on}.
Unlike the $CP^{N-1}$ model, the O($M$) CFT does not have a conserved topological current. For the case of $M=3$, the analog of the
topological current is $\widetilde{B}_\mu = \frac{1}{4} \epsilon_{\mu\nu\lambda} \epsilon_{abc} \phi_a \partial_\nu \phi_b \partial_\lambda \phi_c$.
However, $\widetilde{B}_\mu$ is now {\em not\/} conserved. Physically, this is because the O(3) CFT includes amplitude fluctuations of the $\phi_a$ field,
and so allows spacetime locations with $\phi_a = 0$, which represent `hedgehogs' where Skyrmion number conservation is violated.
Consequently the correlators of $\widetilde{B}_\mu$ do not obey Eq.~(\ref{eq:JJ}) under the O(3) CFT, and are instead characterized by an anomalous dimension
which was computed by Fritz {\em et al.} \cite{fritz}.

Our computations of these current correlators were made possible by a direct evaluation of the Feynman graphs in momentum
space using an algorithm which is described in Appendix~\ref{app:tensoria}. These methods should also 
be applicable to other conserved current correlators of CFTs, including those of the stress-energy tensor \cite{cardy87}, and multi-point 
correlators \cite{suvrat13}.

In Sec.~\ref{sec:ssb}, we will extend our results away from the CFT, into the phase with broken global SU($N$) symmetry of the CP$^{N-1}$ model.
We work out the 
diagrammatic structure that allows divergence-free computation of correlators close to the critical point. 
In particular, we point out the importance of ghost fields in unitary gauge to fulfill Goldstone's theorem. 
We compute the correlation length exponent coming from the symmetry-broken phase and obtain
\bea
\nu = 1-\frac{48}{N \pi^2}\;,
\label{eq:nu}
\eea
in agreement with Halperin {\it et al.'s} computation from 1974 \cite{halperin74} and that of 
the Ekaterinburg group in 1996 \cite{irkhin96}. In the symmetry-broken phase, the sum of logarithmically divergent coefficients 
multiplying the Higgs mass of the gauge propagator yield Eq.~(\ref{eq:nu}).
Finally we note our recent article \cite{huh13_short}, in which we compute the dynamical 
excitation spectrum of the vector boson using the approach of Sec.~\ref{sec:ssb}.

\section{ CP$^{N-1}$ Model}
\label{sec:model}

\begin{figure}[]
\includegraphics[width=40mm]{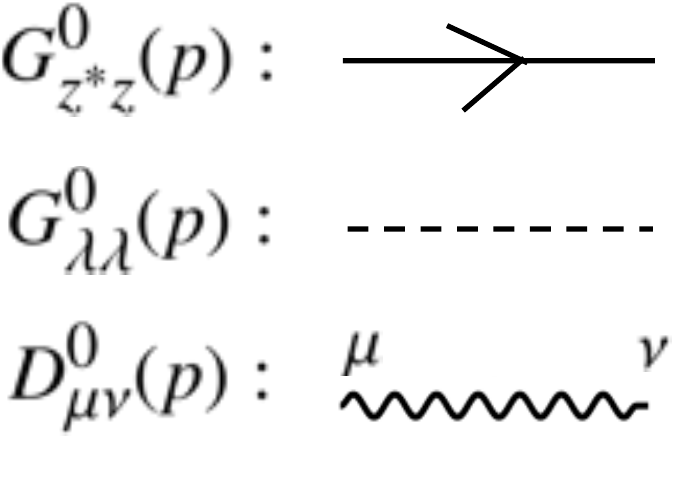}
\;\;\;\;\;
\includegraphics[width=105mm]{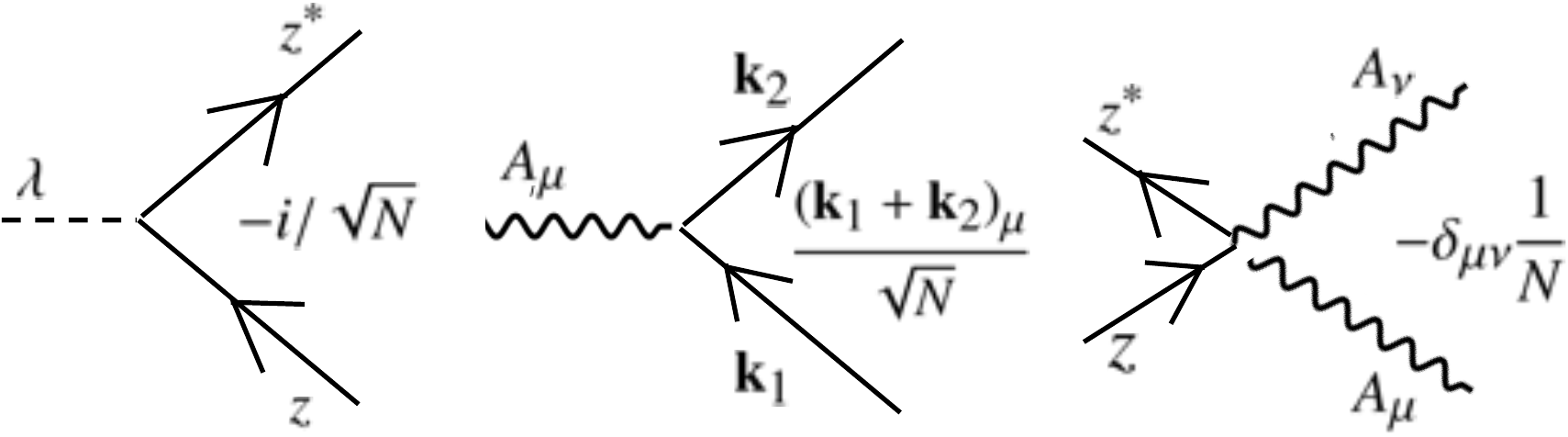}
\caption{Propagators and vertices of $CP^{N-1}$ model used in the $1/N$-expansion.}
\label{fig:3props}
\end{figure}

The CP$^{N-1}$ model describes the dynamics of charged bosons $z^\ast$, $z$ 
minimally coupled to an Abelian gauge field $A_\mu$ \cite{polyakov_book,coleman_book}, with 
subscripts $\mu$, $\nu$ describing coordinates in 2+1 dimensional Euclidean space-time. The 
charged bosons fulfill a unit-length constraint, $\sum_{\alpha=1}^N |z_\alpha|^2=1$,  
at all points in time and space. Upon rescaling $z$ and $A_\mu$ convenient for $1/N$-expansion 
we can write the partition function as 
\beq
Z = \int \mathcal{D} z_\alpha \mathcal{D} \lambda \mathcal{D}  A_\mu \exp \left( -
 \int_\bx \left[ |(\partial_\mu - i A_\mu/\sqrt{N} ) z_a|^2 + i \frac{\lambda}{\sqrt{N}} (|z_a|^2 - N/g)
 \right] \right) \label{Z}\;,
\eeq
where the integration over space and (imaginary) time has been collected in $\int_{\bf x}$. A sum over doubly 
occuring indices is implicit.
The fields have been rescaled such that the relevant coupling constant $g$, which determines the properties and 
phases the model finds itself in, appears in the brackets multiplying the 
(fluctuating) Lagrange multiplier field $\lambda$.  Feynman rules for vertices and propagators are shown in 
Fig.~\ref{fig:3props}.

The large-$N$ expansion is performed by first integrating the $z^\ast$, $z$ and expanding the still dynamical determinant 
to quadratic order in the fields $A_\mu$ and $\lambda$. At $N\rightarrow \infty$ this is exact and the resulting effective action is
\begin{align}
\mathcal{S}_{A-\lambda}=\int_p \frac{ \Pi_{\lambda}(p)}{2} |\lambda(p)|^2 
+
 \frac{\Pi_A(p)}{2}\left(\delta_{\mu\nu}-\frac{p_\mu p_\nu}{p^2}\right)A(-p)_\mu A(p)_\nu
\end{align}
with $\Pi_\lambda(p) = \frac{1}{8p}$ and $\Pi_A(p)=\frac{p}{16}$ \cite{kaul08}. The polarization bubbles of the $\lambda$ and 
$A_\mu$ fields determine the $N\rightarrow \infty$ propagators:
\begin{align}
\langle \lambda(-p)\lambda(p)\rangle_{N\rightarrow \infty} &= G^0_{\lambda\lambda}(p)= 8 p 
\nonumber\\
\langle A_\mu(-p)A_\nu(p)\rangle_{N\rightarrow \infty}& = D^0_{\mu\nu}(p) = \frac{16}{p}
\left(\delta_{\mu\nu}-\frac{p_\mu p_\nu}{p^2}\right)
\nonumber\\
\langle z^\ast(p) z(p) \rangle &= G^0_{z^\ast z}(p) = \frac{1}{p^2}\;.
\label{eq:1loop}
\end{align}
Note that for the gauge field propagator, there is another diagram with a $z$ loop attached by a $z^\ast zAA$ vertex to the gauge field propagator. However, this is conventionally dropped as it would be zero in 
dimensional regularization \cite{kaul08}. Without loss of generality for physical observables, 
we will use (transversal) Landau gauge.  

\subsection{Relation to spin observables of deconfined quantum magnets}
\label{subsec:spin}

We here recapitulate how the emerging quantum fields $z^\ast$, $z$, and $A_\mu$ are related to spin observables in the deconfined critical theory \cite{japan05}. 
First of all, the ``original'' unit-length N\'eel spin vector field $\mathbf{n}(\mathbf{x})$ can be parametrized as a bilinear of complex-valued spinon fields $z^\ast_\alpha$,
\begin{align}
\mathbf{n}(\mathbf{x}) = z^\ast_\alpha(\mathbf{x}) \boldsymbol{\sigma}_{\alpha\beta} z_\beta(\mathbf{x})\;,
\label{eq:CP_1}
\end{align}
where for $S\!U(2)$-spins, $\boldsymbol{\sigma}$ is a vector of Pauli matrices and the 
``flavor'' indices $\alpha$, $\beta$ run over $1$ and $2$. The scaling dimension of the 
N\'eel ordering field has been computed for the CP$^{N-1}$ model in Ref.~\onlinecite{kaul08}.
The local $U(1)$ transformation $z(\mathbf{x}) \rightarrow z(\mathbf{x})  e^{i \theta(\mathbf{x})}$ leaves the Lagrangian and $\mathbf{n}(x)$ invariant, and requires a $U(1)$ gauge field, $A_\mu$. We can further relate spin observes to the gauge field $A_\mu$:
the staggered vector spin chirality can be written as, 
\beq
B_\mu = \epsilon_{\mu\nu\lambda} \partial_\nu A_\lambda = {\textstyle \frac{1}{4}} \epsilon_{\mu\nu\lambda} \mathbf{n} \cdot \left( \partial_\nu \mathbf{n} \times
\partial_\lambda \mathbf{n} \right)\;. \label{chirality}
\eeq
The last term specifies how $B_\mu$ can be related to the operators of the underlying antiferromagnet,
and identifies it as the Skyrmion current: the spatial integral of its temporal component $B_t$ is the Skyrmion number of the
texture of the N\'eel order parameter underlining the topological nature of the vector boson.

The corresponding vector chirality operator for the $O(3)$ model is 
\beq
\widetilde{B}_{\mu} = \frac{1}{2} \epsilon_{\mu\nu\lambda} \epsilon_{abc} \phi_a \partial_\nu \phi_b \partial_\lambda \phi_c  \;,
\eeq
which is the analog of the flux operator for the confining critical point described by the $\phi^4$ field theory of the 3-component field $\phi_a$. Indeed, such correlations were measured recently by Fritz {\em et al.} \cite{fritz} in quantum Monte Carlo. 
This quantity can also be measured in Raman scattering \cite{ss,nl} if the light couples preferentially to one sublattice
of the antiferromagnet. 

Thus knowing the gauge field properties will allow us to compare the vector spin chirality of the $CP^{1}$ model versus 
that of the $O(3)$ model. 

\section{Universal magnetic transport: current correlator $\langle J_\mu J_\nu \rangle$}
\label{sec:critical_point}

In this section, we compute the current-current correlator of the $CP^{N-1}$ model to order $1/N$. As a warm up, and 
to make contact with previous work, we compute the same quantity for the $O(M)$ vector 
to order $1/M$. This is important for two reasons: (i) to make predictions for a variety of physical situations these field theories are believed to describe and (ii) to classify interacting CFT$_3$'s by means of 
numerical constants that determine their correlation functions.

\subsection{Warm-up: Universal conductance $C_J$ of the $O(M)$-model to $1/M$}
\label{subsec:cj_on}

\begin{figure}[b]
\includegraphics[width=60mm]{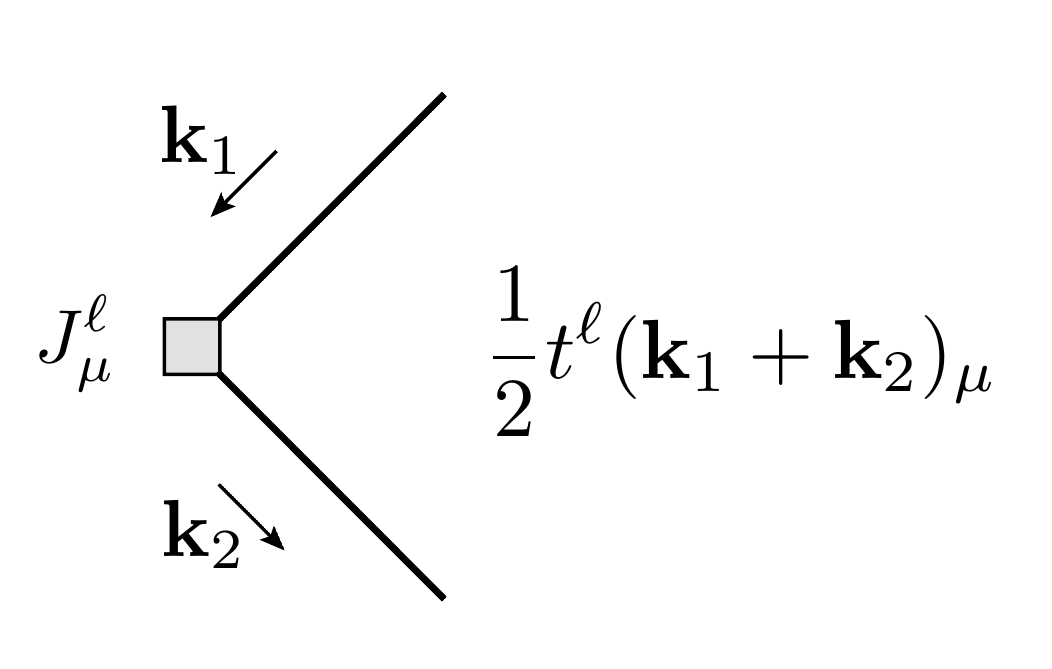}
\caption{Feynman rule for the $O(M)$ current vertex. Here the straight line is the propagator for 
real fields $\langle \phi_a (-p) \phi_b(p) \rangle = \delta_{ab}/p^2$. }
\label{fig:feyn_on_current}
\end{figure}

In absence of the gauge field $A_\mu$, the critical behavior of Eq.~(\ref{Z}) is equivalent to that of the 
$O(M)$ vector model with a quartic self-interaction. The latter arises 
as the most relevant term upon ``softening'' the unit-length constraint into a series of polynomial interaction terms.
We use the conventions and action following Ref.~\onlinecite{podolsky1} which is reproduced here. 
\begin{align}
\mathcal{S}_{O(M)} = \frac{1}{2} \int_{\bx}   \left[ (\partial \phi_\alpha)^2 + \frac{i}{\sqrt{M}} \lambda \left( \phi_\alpha^2 - M/g \right) + \frac{\lambda^2}{4u} \right]
\label{eq:on_action}
\end{align}
The conserved current in the vector $O(M)$-model of $M$-component real fields $\phi_a$ is
\begin{align}
J^\ell_\mu = -i \left(\phi_a t^{\ell}_{ab} \partial_\mu \phi_b\right) 
 = \frac{1}{2} (\bk_1+\bk_2)_\mu \phi_a t^{\ell}_{ab} \phi_b \;
\label{eq:def_on_current}
\end{align}
where the generators $t^\ell$ are purely imaginary antisymmetric matrices conventionally normalized as 
Tr$(t^\ell t^m)=2 \delta^{\ell m}$. The second equality notes the symmetrized version in momentum space, where $\bk_1$ and $\bk_2$ are the incoming and outgoing momenta from the current vertex. This version of the $O(M)$ current vertex definition is exhibited in Fig.~\ref{fig:feyn_on_current}.

The $M=3$ case for magnets is our primary interest in the present paper. 
$M=2$ corresponds to the XY-universality class of the $O(2)$ model in 2+1 dimensions, 
believed to describe the superfluid-to-insulator phase transition in ultracold atoms and 
superconducting films. The conserved current in Eq.~(\ref{eq:def_on_current}) is then 
associated with a conserved $U(1)$ global charge symmetry. For the electrically neutral ultracold 
atoms, this is simply the conservation of particles or ``number charge''. For superconducting films, 
the bosons carry electric charge (twice that of the constituent electrons) and the application of the 
Kubo formula to $\langle J J \rangle$ actually yields the universal, electrical DC-conductivity applicable 
to the regime $\omega \gg k_B T$ as mentioned in the introduction.

This universal high-frequency conductance of the $O(2)$ model at the critical point 
was computed in $\epsilon$-expansion by Fazio and Zappala \cite{fazio96} and in $1/M$-expansion by 
Cha {\it et al.} \cite{cha91}. At the critical point, the $O(M)$ model becomes a strongly interacting 
conformal field theory in 2+1 dimensions (CFT$_3$)
and two-point correlators of conserved quantities are constrained by conformal symmetries 
\cite{cardy87,osborn94,petkou95,petkou96,erdmenger97,maldacena11,bzowski12,bzowski13}. 
The current-current correlator 
is determined by a single parameter $C_J$
\begin{align}
\langle J^\ell_\mu(-p) J^m_\nu(p) \rangle = -C_J \delta^{\ell m}\, |\bp| \left(\delta_{\mu\nu}-\frac{\bp_\mu \bp_\nu}{p^2}\right)\;.
\end{align}
In 1991, Cha {\it et al.} \cite{cha91} reported a value for $C_J$ to order $1/M$ in momentum space. We will below obtain the same value for $C_J$ within our 
momentum space computation using a newly developed algorithm to evaluate tensor-valued 
momentum integrals (see Appendix \ref{app:tensoria} for details). 
The advantage of this approach is that 
it is straightforward to generalize to more complicated situations, such as gauge theories. 

\begin{figure}[b]
\vspace{5mm}
\includegraphics[width=120mm]{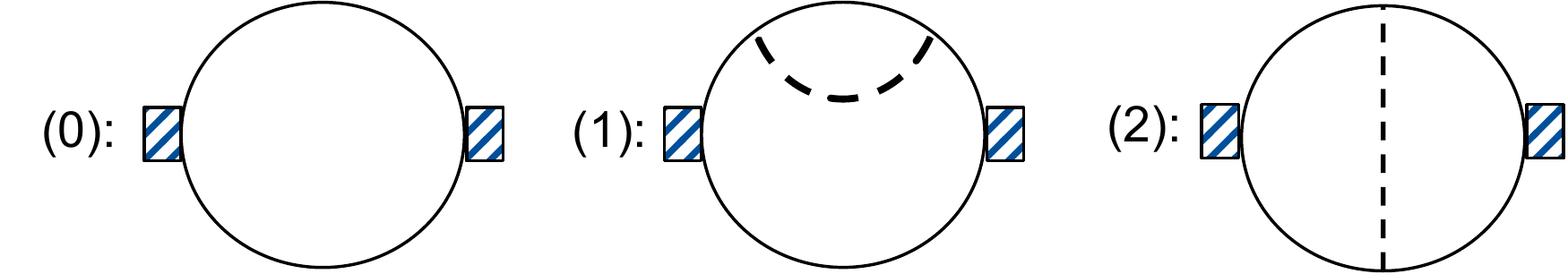}
\caption{Diagrams for the current-current correlator in the $O(M)$ model. (0) is the only diagram at $M\rightarrow \infty$.
(1) and (2) are of order $1/M$. Blue-dashed boxes are $O(M)$ current-vertices, where the current is 
defined in Eq.~(\ref{eq:def_on_current}). Dashed line is the $\lambda$ propagator
$\langle \lambda(-p)\lambda(p)\rangle^{O(M)}_{M\rightarrow \infty} =  16 p$
 which, for real fields, comes with an additional factor of 2 compared to Eq.~(\ref{eq:1loop}) \cite{podolsky1}.}
\label{fig:current_on}
\end{figure}

The diagrammatic evaluation of the current two-point function 
now proceeds as follows: one connects two current insertions in all possible ways using the propagators in 
Fig.~\ref{fig:3props}, counting factors of $M$. At $M\rightarrow\infty$, this yields the bubble 
diagram (0) in Fig.~\ref{fig:current_on}. At order $1/M$, one obtains (1) and (2) in the same figure. 
Note that diagrams involving closed loops with an odd number of current insertions vanish by symmetry as the
trace over a single generator $t^\ell$ is zero.
The expressions for these three diagrams are 
\begin{alignat}{2}
J_{\mu\nu}^{\ell m}  (p)^{(0)} &=  \text{Tr}_{O(M)}
\Bigg[
\int_{\mathbf{k}}
 \frac{t^\ell (2\bk+\bp)_\mu t^m (2\bk+\bp)_\nu }{ 4 k^2 (\bk+\bp)^2 } \Bigg]
  & a_0=2\nn
J_{\mu\nu}^{\ell m}  (p)^{(1)} &= 
\text{Tr}_{O(M)}
\Bigg[
\int_\bq \int_\bk
 \frac{t^\ell (2\bk+\bp)_\mu t^m (2\bk+\bp)_\nu }{ 4 k^4 (\bk+\bp)^2 }
 \left(
 \frac{-i}{\sqrt{M}}
 \right)^2 16 q
 \left(
 \frac{1}{\left(\bk + \bq\right)^2}- \frac{1}{q^2}
 \right)
 \Bigg]
~~~~& a_1=4\nn
J_{\mu\nu}^{\ell m}  (p)^{(2)} &= 
\text{Tr}_{O(M)}
\Bigg[
\int_\bq \int_\bk
 \frac{t^\ell (2\bk+\bp)_\mu t^m (2(\bk+\bq)+\bp)_\nu }{ 4 k^2 (\bk+\bp)^2 (\bk+\bq)^2 (\bk+\bp +\bq)^2  }
 \left(
 \frac{-i}{\sqrt{M}}
 \right)^2 16 q
 \Bigg]
 \;\;~~~~~& a_2=2
\label{eq:current_on}
\end{alignat}
with abbreviated 3d momentum integrations $\int_{\mathbf{k}}= \int \frac{ d^3 \bk}{8\pi^3}$.
In diagram (1), we have subtracted the zero-momentum ($\bk=0$) value of the self-energy insertion 
on the internal propagator in order to operate at the renormalized position of the critical point.
We will write the current correlator as a momentum-decomposed sum of its diagrammatic contributions
\begin{align}
\langle J^\ell_\mu(-p) J^m_\nu(p) \rangle 
&= \delta^{\ell m}\sum_{i = 0}^{2}  a_i J^{(i)} _{\mu\nu}  (p)
\equiv 
\sum_{i=0}^2 a_i \left[ I^{(i)}_{\text{T}}(p)\left(\delta_{\mu\nu}-\frac{\bp_\mu \bp_\nu}{p^2}\right) 
+
I^{(i)}_{\text{L}}(p) \frac{\bp_\mu \bp_\nu}{p^2} 
\right]\;.
\label{eq:JJ_def}
\end{align}
The momentum index structure, in general, can be decomposed into transversal, longitudinal and odd parts. 
With Chern-Simons terms, there may also be odd parts, but they are immaterial for our subsequent discussion. 
We evaluate the tensor-valued momentum integrals in Eq.~(\ref{eq:current_on}) using Tensoria and subsequently decompose them into transversal and longitudinal components as shown in 
Table \ref{tab:c_j_on}.
\begin{table} 
\centering
\begin{tabular}{ccccc}
\text{Diagram } &~$ I^{(i)}_{T}(p)$		~~~~&  $I^{(i)}_{\text{L}}(p)$ &  Log-Singularity (transverse) & Factor  \\[1mm]
\hline
0 & $-\frac{1}{32} p$                               & 0                                                 &       0                                                                          & 2   \\[1mm]
1 & $\frac{1}{M}\frac{13}{144\pi^2} p $   & $ \frac{1}{M}\frac{1}{24\pi^2} p$ &$\frac{1}{M}\frac{1}{12\pi^2} p\log{\frac{\Lambda}{p}}$& 4  \\[1mm]
2 & $\frac{1}{M}\frac{1}{24\pi^2}     p $   & $-\frac{1}{M}\frac{1}{12\pi^2} p$ &$-\frac{1}{M}\frac{1}{6\pi^2}   p\log{\frac{\Lambda}{p}} $& 2  \\[1mm]
\hline
\end{tabular}
\caption{Evaluated contributions to $C_J$ at the critical point for the $O(M)$ model. The longitudinal components 
add to 0 after accounting for the diagram factors $a_i$'s in Eq.~(\ref{eq:JJ_def}). The transverse components give 
Eq.~(\ref{eq:value_cj_on}). The fourth column shows the mutually canceling log-singularities that appear in the 
transverse parts of the individual diagrams.}
\label{tab:c_j_on}
\end{table}
As we explain in more depth in Appendix \ref{app:tensoria}, Tensoria is built around recursion relations 
of Davydychev \cite{davy91,davy92,bzowski12,suvrat13} that transform tensor-valued momentum integrals 
into a permuted series of scalar integrals.
Adding the values in Table \ref{tab:c_j_on}, we obtain
\begin{align}
C^{O(M)}_J=\frac{1}{16}\left(1-\frac{1}{M} \frac{64}{9\pi^2}\right)
\approx \frac{1}{16}\left(1-\frac{1}{M} 0.72\right)
\label{eq:value_cj_on}
\end{align}
in agreement with Cha {\it et al.}\cite{cha91}. We see that here the large-$M$ expansion 
works satisfactorily down to $M=3$, where the correction to the 
$M\rightarrow \infty$ value is $\sim 25\%$.

It is a strong check on our algorithm 
that all logarithmic singularities in the fourth column of Table \ref{tab:c_j_on} cancel out. For non-conserved operators, 
such log-singularities as a function of momentum generate anomalous scaling dimensions at criticality. Because the 
$1/M$-expansion fulfills Ward identities between self-energy and vertex corrections (diagrams (1) and (2)), we correctly 
recover the required result that no anomalous dimension is generated for the conserved current \cite{gross75,franz03}.

\subsection{$C_J$ of the CP$^{N-1}$ model to $1/N$}
\label{subsec:cj_cpn}
\begin{figure}[b]
\includegraphics[width=100mm]{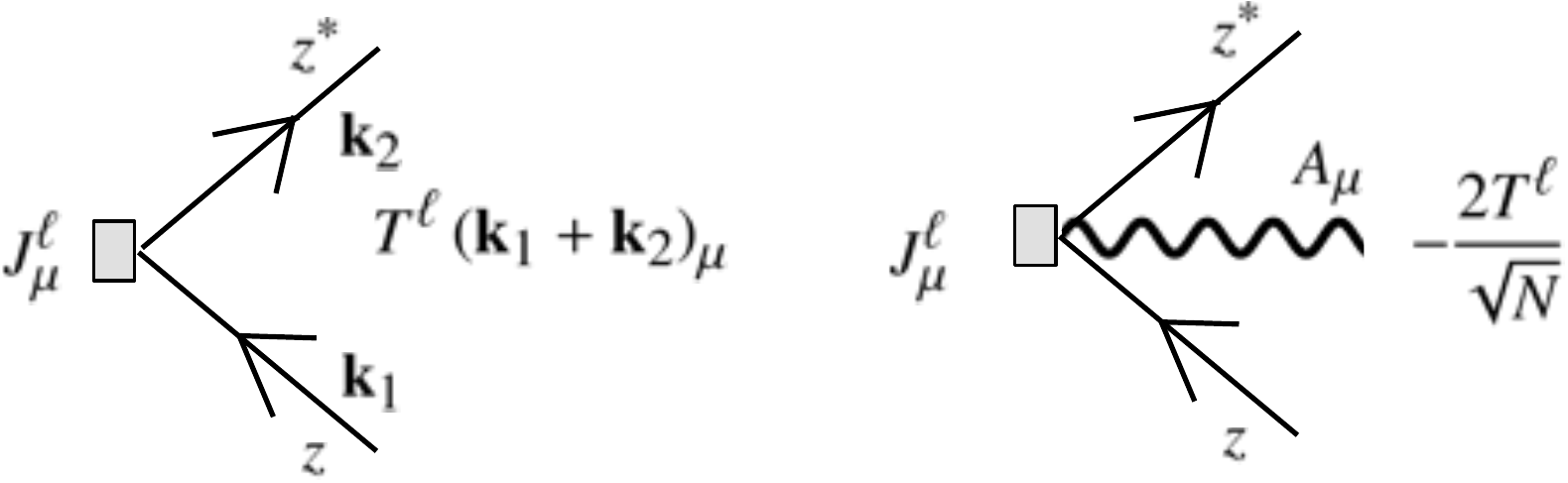}
\caption{Feynman rules for the $S\!U(N)$ flavor current vertices.}
\label{fig:feyn_current}
\end{figure}

We now compute the corresponding current-current correlator of the $CP^{N-1}$ model. With superscript $\ell$ as the generator index, subscripts $a$ and $b$ as component indices in flavor space, and subscript $i$ as the spatial direction index, the $S\!U(N)$ flavor current is  
\begin{align}
J^\ell_{i}=-i z_a^\ast T^\ell_{ab}\left(D_i z_b\right) + i \left(D_i z_a\right)^\ast T^\ell_{ab} z_b\;,
\label{eq:flavor_current}
\end{align}
using the covariant derivative $D_i=\partial_i-iA_i/(\sqrt{N})$. Here, $T^\ell$'s are generators of the $S\!U(N)$ group normalized 
so that Tr$(T^\ell T^m)=\delta^{\ell m}$. The Feynman rules for the flavor current vertices are given in Fig.~\ref{fig:feyn_current}.
In the context of quantum magnets at the critical point (deconfined with gauge field and conventional without the gauge field), this current can be related to the magnetization \cite{subir94}.

\begin{figure}[t]
\includegraphics[width=90mm]{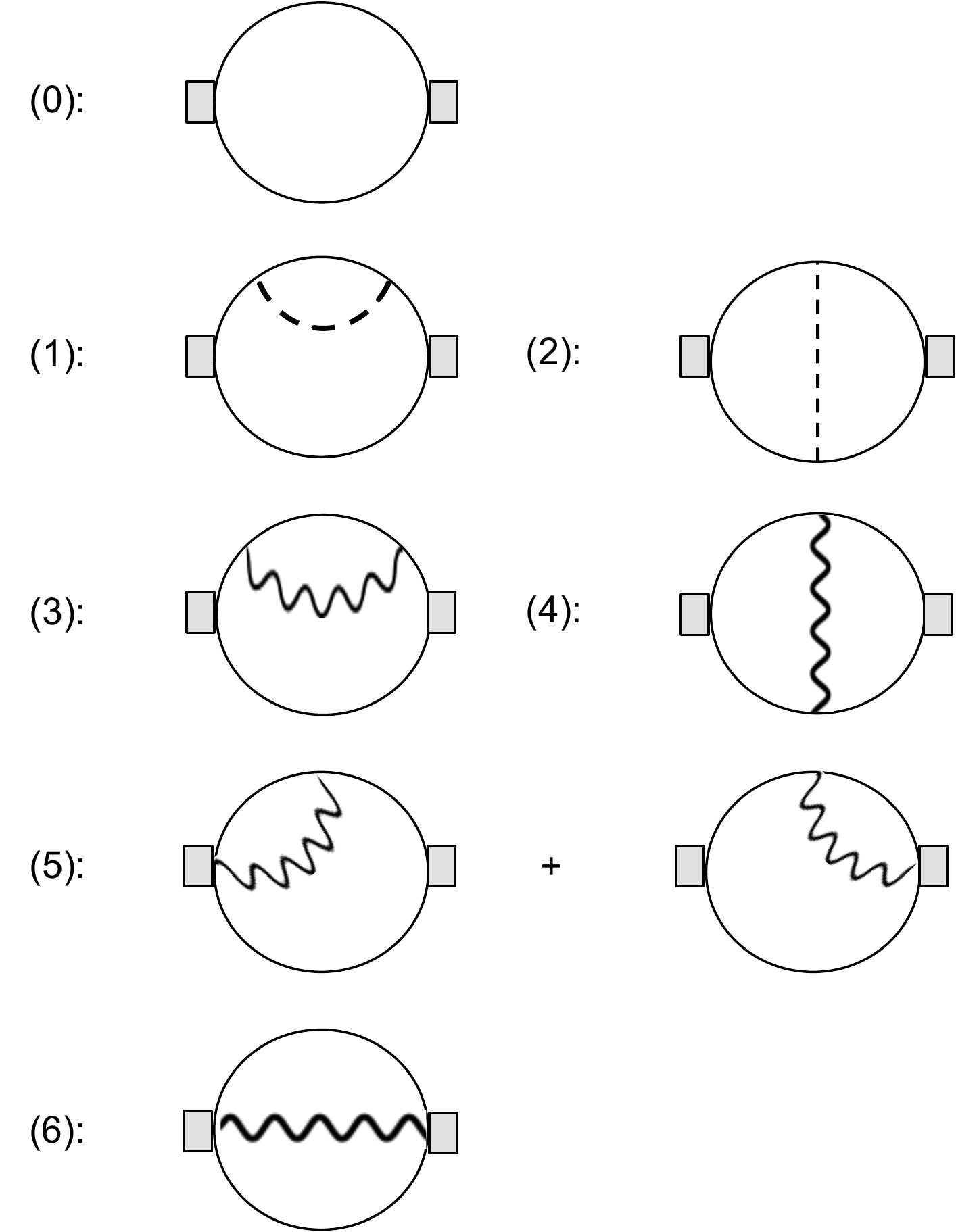}\\[-5mm]
\caption{Diagrams for the current-current correlator of the CP$^{N-1}$ model to order $1/N$.}
\label{fig:JJ}
\end{figure}
The diagrams to evaluate are exhibited in Fig.~\ref{fig:JJ}. Note that in diagram (3), the self-energy insertion of the gauge field does not 
renormalize the position of the critical point and we therefore do not have to subtract the zero-momentum value here.
We can safely ignore the self-energy insertion of the gauge-field loop on the internal $z$-propagator connected 
via a four-point vertex $z^\ast z A A$ (as in Fig.~\ref{fig:orderpi} (b)) as this can be absorbed into a shifted critical point. 
The analytic expressions are
\begin{alignat}{2}
J_{\mu\nu}^{\ell m}  (p)^{(0)} &=
\text{Tr}_{S\!U(N)}
\Bigg[
\int_{\mathbf{k}}
 \frac{T^\ell (2\bk+\bp)_\mu T^m (2\bk+\bp)_\nu }{ k^2 (\bk+\bp)^2 } \Bigg]
  & a_0=1\nn
J_{\mu\nu}^{\ell m}  (p)^{(1)} &= 
\text{Tr}_{S\!U(N)}
\Bigg[
\int_\bq \int_\bk
 \frac{T^\ell (2\bk+\bp)_\mu T^m (2\bk+\bp)_\nu }{ k^4 (\bk+\bp)^2 }
 \left(
 \frac{-i}{\sqrt{N}}
 \right)^2 8 q
 \left(
 \frac{1}{\left(\bk + \bq\right)^2}- \frac{1}{q^2}
 \right)
 \Bigg]
~~~~& a_1=2\nn
J_{\mu\nu}^{\ell m} (p)^{(2)} &= 
\text{Tr}_{S\!U(N)}
\Bigg[
\int_\bq \int_\bk
 \frac{T^\ell (2\bk+\bp)_\mu T^m (2(\bk+\bq)+\bp)_\nu }{ k^2 (\bk+\bp)^2 (\bk+\bq)^2 (\bk+\bp +\bq)^2  }
 \left(
 \frac{-i}{\sqrt{N}}
 \right)^2 8 q
 \Bigg]
 \;\;~~~~~& a_2=1
 \nn
J^{\ell m}_{\mu\nu} (p)^{(3)} &= 
\text{Tr}_{S\!U(N)}
\Bigg[
\int_\bq \int_\bk 
   \frac{(2 \bk+\bp)_\mu T^\ell (2\bk+\bp)_\nu T^m  }{ k^4 (\bp+\bk)^2}
 \frac{ (2\bk+\bq)_\lambda (2\bk+\bq)_\rho }{  (\bk+\bq)^2 }   
  \left(\frac{ \delta_{\lambda\rho} q^2 - q_\lambda q_\rho}{ q^3} \right) 
  \frac{16}{N} 
  \Bigg]
   & ~~~~~a_3=2\nn
J^{\ell m}_{\mu\nu} (p)^{(4)} &=
 \text{Tr}_{S\!U(N)}
 \Bigg[
 \int_\bq \int_\bk 
   \frac{(2 \bk+\bp)_\mu T^\ell (2 (\bk + \bq) +\bp)_\nu T^m (2 (\bk + \bp)+\bq)_\lambda (2\bk+\bq)_\rho }{ k^2 (\bp+\bk)^2 (\bk+\bp+\bq)^2 (\bk+\bq)^2}
  \left(\frac{ \delta_{\lambda\rho} q^2 - q_\lambda q_\rho}{ q^3} \right) 
   \frac{16}{N} 
   \Bigg]  
  & ~~a_4=1
  \nn
J^{\ell m}_{\mu\nu} (p)^{(5)} &= 
 \text{Tr}_{S\!U(N)}
 \Bigg[
 \int_\bq \int_\bk 
  \delta_{\mu\lambda} (-2 T^\ell)
  \frac{(2\bk+\bq)_\rho (2\bk+\bp)_\nu T^m }{ k^2 (\bk+\bq)^2 (\bk+\bp)^2 } 
  \left(\frac{ \delta_{\lambda\rho} q^2 - q_\lambda q_\rho}{ q^3} \right) 
  \frac{16}{N}
+ (\mu \leftrightarrow \nu, \ell \leftrightarrow m)
  \Bigg]
  & a_5 =2 \nn
J^{\ell m}_{\mu\nu} (p)^{(6)} &=\text{Tr}_{S\!U(N)}
\Bigg[
\int_\bq \int_\bk 
\delta_{\mu\lambda} (-2T^\ell)
  \frac{1}{ (\bk+\bq)^2 ( \bk + \bp )^2  }  
 \left(\frac{ \delta_{\lambda\rho} q^2 - q_\lambda q_\rho}{ q^3} \right) 
   \delta_{\nu\rho}(-2 T^m)
    \frac{16}{N} 
  \Bigg]
  & a_6=1 \nn
 \label{eq:current_gauge_on}
\end{alignat}

\begin{table}[ht!]
\centering
\begin{tabular}{ccccc}
\text{Diagram} &~ $I_T^{(i)}(p)$ ~~~& $I_L^{(i)}(p)$ & Log-Singularity (transverse) & \text{Factor} \\[1mm]
\hline
0 & $-\frac{1}{16} p$                               & 0                                                 &       0                                                                          & 1   \\[1mm]
1 & $\frac{p}{N}\frac{13}{144\pi^2} $   & $ \frac{p}{N}\frac{1}{24\pi^2} $ &$\frac{p}{N}\frac{1}{12\pi^2} \log{\frac{\Lambda}{p}}$& 2  \\[1mm]
2 & $\frac{p}{N}\frac{1}{24\pi^2}      $   & $-\frac{p}{N}\frac{1}{12\pi^2} $ &$-\frac{p}{N}\frac{1}{6\pi^2}   \log{\frac{\Lambda}{p}} $&1  \\[1mm]
3 & $\frac{p}{N}\frac{1}{3\pi^2}$           & $-\frac{p}{N} \frac{2}{3\pi^2}$   &$-\frac{p}{N}\frac{4}{3\pi^2}\log{\frac{\Lambda}{p}}  $ &  2   \\[1mm]
4 & $-0.110 \frac{p}{N}		      $            & $-\frac{p}{N} \frac{2}{3\pi^2}$   &0&  1   \\[1mm]
5 & $\frac{p}{N} \frac{13}{9\pi^2}$      & $\frac{p}{N} \frac{2}{\pi^2}$      &$\frac{p}{N} \frac{4}{3\pi^2}\log{\frac{\Lambda}{p}}  $  &  2    \\[1mm]
6 & $-\frac{p}{N} \frac{1}{\pi^2}$        & $-\frac{p}{N} \frac{2}{\pi^2}$      &0&  1   \\[1mm]
\hline
\end{tabular}
\caption{Evaluated contributions to the current-current correlator of the $CP^{N-1}$ model. 
The longitudinal components add to 0 and transverse parts give Eq.~(\ref{eq:value_cj_cp}). 
The log-singularities mutually cancel.}
\label{tab:c_j}
\end{table}
Adding all 7 diagrams from Table \ref{tab:c_j}, we obtain our new value for the flavor current correlator of the CP$^{N-1}$ model
\begin{align}
C^{CP^{N-1}}_J=\frac{1}{16} -\frac{0.171}{N} 
\approx \frac{1}{16}\left(1-\frac{1}{N} \frac{243.67}{9\pi^2}\right)
\approx \frac{1}{16}\left(1-\frac{1}{N} 2.74\right)
\;.
\label{eq:value_cj_cp}
\end{align}
Comparing this to Eq.~(\ref{eq:value_cj_on}), the leading $1/N$ correction relative to the $N\rightarrow \infty$ value is much larger for the CP$^{N-1}$ than the $O(M)$ model: for $N=2$, the correction is larger than the leading term. Relatively large $1/N$ corrections for critical exponents of 
the CP$^{N-1}$ model were also found by Irkhin {\it et al.} \cite{irkhin96}. It would be interesting to compare Eq.~(\ref{eq:value_cj_cp})
to large-scale numerical simulations or conformal field theory methods in position space.

\section{Emergent gauge excitations: vector boson correlator $\langle A_\mu A_\nu \rangle$}
\label{sec:ca}

We now compute the gauge propagator to $1/N$ at the critical point. From the perspective of quantum spin 
systems, detecting the scaling properties of the vector boson in numerical simulations, and ultimately in experiments, 
via the observables discussed in Subsec.~\ref{subsec:spin}, 
would be a signature of an underlying deconfined quantum critical point \cite{huh13_short}. The universal amplitudes contained 
in $\langle A_\mu A_\nu \rangle$ translate to those of the topological current and are of general 
interest to the classification of strongly interacting CFT$_3$'s.

\subsection{$C_A$ of the CP$^{N-1}$ model to $1/N$}

\begin{figure}[]
\includegraphics[width=100mm]{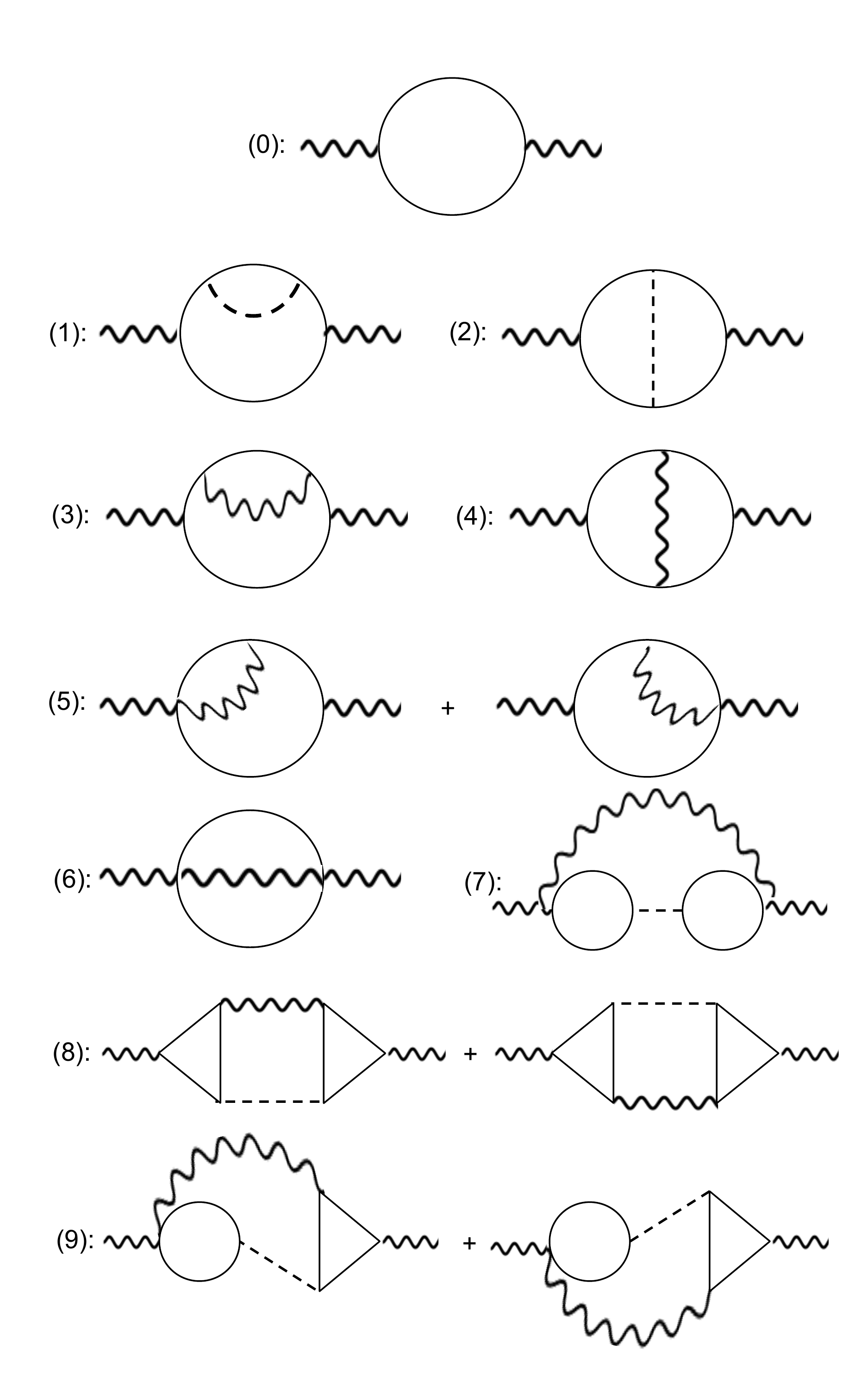}
\caption{Diagrams for the gauge correlator of the $CP^{N-1}$ model at the critical point to order $1/N$.}
\label{fig:diags_gauge_crit}
\end{figure}

The evaluation of Wick's theorem now progresses as before for the current correlator, except that now the 
external legs are gauge fields and not current vertices. The relevant diagrams to evaluate are 
shown in Fig.~\ref{fig:diags_gauge_crit}.
The expressions are given by
\begin{alignat}{2}
\Sigma^{(0)}_{\mu\nu}(p) &= 
\int_{\mathbf{k}}
 \frac{ (2\bk+\bp)_\mu (2\bk+\bp)_\nu }{ k^2 (\bk+\bp)^2 }& a_0 = 1\nn
\Sigma^{(1)}_{\mu\nu} (p) &= -\frac{8}{N} \int_\bq \int_\bk 
   \frac{(2\bk+\bp)_\mu (2\bk+\bp)_\nu\, q }{ k^4 (\bp+\bk)^2 } \left( \frac{1}{(\bk+\bq)^2} - \frac{1}{q^2}\right) 
  & a_1=2\nn
\Sigma^{(2)}_{\mu\nu} (p) &= -\frac{8}{N} \int_\bq \int_\bk 
   \frac{(2\bk+\bp)_\mu ( 2(\bk+\bq)+\bp)_\nu\, q }{ k^2 (\bp+\bk)^2 (\bk+\bq)^2 (\bp+\bk+\bq)^2 } 
   & a_2=1\nn
\Sigma^{(3)}_{\mu\nu} (p) &= \frac{16}{N} \int_\bq \int_\bk 
   \frac{(2 \bk+\bp)_\mu (2\bk+\bp)_\nu  }{ k^4 (\bp+\bk)^2}
   \left( \frac{ (2\bk+\bq)_\lambda (2\bk+\bq)_\rho }{  (\bk+\bq)^2 }  -  \frac{ q_\lambda q_\rho }{q^2} \right) 
   \left(\frac{ \delta_{\lambda\rho} q^2 -  q_\lambda q_\rho}{q^3} \right) 
   & a_3=2\nn
\Sigma^{(4)}_{\mu\nu} (p) &= \frac{16}{N} \int_\bq \int_\bk 
   \frac{(2 \bk+\bp)_\mu (2 (\bk + \bq) +\bp)_\nu (2 (\bk + \bp)+\bq)_\lambda (2\bk+\bq)_\rho }{ k^2 (\bp+\bk)^2 (\bk+\bp+\bq)^2 (\bk+\bq)^2 }
      \left(\frac{ \delta_{\lambda\rho} q^2 -  q_\lambda q_\rho}{q^3} \right) 
   & ~~a_4=1\nn
\Sigma^{(5)}_{\mu\nu} (p) &= -\frac{16}{N} \int_\bq \int_\bk 
  \frac{(2\bk+\bq)_\rho (2\bk+\bp)_\nu }{ k^2 (\bk+\bq)^2 (\bk+\bp)^2 } 
  \left(\frac{ \delta_{\lambda\rho} q^2 -  q_\lambda q_\rho}{q^3} \right) 
 \delta_{\mu\lambda}
 + (\mu \leftrightarrow \nu)  
  & a_5 =4 \nn
\Sigma^{(6)}_{\mu\nu} (p) &= \frac{16}{N} \int_\bq \int_\bk 
  \frac{1}{ (\bk+\bq)^2 ( \bk + \bp )^2 }  
   \left(\frac{ \delta_{\lambda\rho} q^2 -  q_\lambda q_\rho}{q^3} \right) 
  \delta_{\mu\lambda} \delta_{\nu\rho}
  & a_6=4 \nn
\Sigma^{(7)}_{\mu\nu} (p) &= -\frac{128}{N} \int_{\bq,\bk,\bl}  
  \frac{1}{k^2(\bk+\bp+\bq)^2}\frac{1}{l^2(\bl+\bp+\bq)^2}
  |\bp+\bq|
     \left(\frac{ \delta_{\lambda\rho} q^2 -  q_\lambda q_\rho}{q^3} \right) 
  & a_{7}=4 \nn
\Sigma^{(8)}_{\mu\nu} (p) &= -\frac{128}{N}  \int_{\bq,\bk,\bl} \Bigg[ 
  \frac{(2\bk+\bp)_\mu (2\bk+\bq)_\lambda }{ k^2 (\bk+\bp)^2 (\bk+\bq)^2} 
  \frac{(2\bl+\bq)_\rho (2\bl+\bp)_\nu }{ l^2 (\bl+\bp)^2 (\bl+\bq)^2} 
 |\bp-\bq|
    \left(\frac{ \delta_{\lambda\rho} q^2 -  q_\lambda q_\rho}{q^3} \right) 
\nn
  &+\frac{(2\bk+\bp)_\mu (2\bk+\bp+\bq)_\lambda }{ k^2 (\bk+\bp)^2 (\bk+\bq)^2} 
  \frac{(2\bl+\bp+\bq)_\rho (2\bl+\bp)_\nu }{ l^2 (\bl+\bp)^2 (\bl+\bq)^2} 
  q
  \left(\frac{ \delta_{\lambda\rho} |\bp-\bq |^2 -  (\bp-\bq)_\lambda (\bp-\bq)_\rho}{|\bp-\bq |^3} \right) 
  \Bigg]
   & a_{8}=2 \nn
\Sigma^{(9)}_{\mu\nu} (p) &= \frac{128}{N} \int_{\bq,\bk,\bl}
 \frac{ \delta_{\mu\lambda}}{k^2(\bk+\bp+\bq)^2} \left( \frac{(2\bl-\bq)_\rho(2\bl+\bp)_\nu}{l^2(\bl-\bq)^2(\bl+\bp)^2} 
  +  \frac{(2\bl+2\bp+\bq)_\rho(2\bl+\bp)_\nu}{l^2(\bl+\bp)^2(\bl+\bp+\bq)^2}\right)
  |\bp+\bq|
    \left(\frac{ \delta_{\lambda\rho} q^2 -  q_\lambda q_\rho}{q^3} \right) 
  & a_{9}=4 \nn
  \label{eq:gauge_crit}
\end{alignat}
Note that in these expressions, the trace over $N$ flavor components is included in counting factors of $N$. We evaluate these diagrams in momentum space using Tensoria (cf. App.~\ref{app:tensoria}). The renormalized gauge propagator remains
transverse at the critical point in accordance with symmetry constraints. As before, all the log-singularities that depend on  
momenta cancel as summarized in Table \ref{tab:aa_crit}. 
It is a strong check on the expansion and calculation procedures that we explicitly see these log-singularities cancel.

The renormalized form of the gauge propagator thus becomes
\begin{align}
D_{\mu\nu}(p) = \frac{1}{\left[D^0_{\mu\nu}(p)\right]^{-1} - \Sigma_{\mu\nu}(p)} \;,
\end{align}
where the $N\rightarrow \infty$ value is $D^0_{\mu\nu}(p)$ and the $1/N$ corrections are $\Sigma_{\mu\nu}(p)=\sum_{i=1}^9 a_i \Sigma^{(i)}_{\mu\nu}(p)$.
As before, we split the self-energy into transversal and longitudinal parts
\begin{align}
\Sigma_{\mu\nu}(p)=\sum_{i=1}^9 a_i \left[
\Sigma^{(i)}_{\text{T}}(p)\left(\delta_{\mu\nu}-\frac{\bp_\mu \bp_\nu}{p^2}\right) 
+
\Sigma^{(i)}_{\text{L}}(p) \frac{\bp_\mu \bp_\nu}{p^2}
\right]
+
\frac{8\Lambda}{N\pi^2}\delta_{\mu\nu}
\;.
\end{align}
The last term proportional to $\Lambda$ can be safely absorbed into the location of the critical point. 
\begin{table} [t]
\centering
\begin{tabular}{ccccc }
\text{Diagram} &~ $\Sigma_T^{(i)}(p)$ ~~~& $\Sigma_L^{(i)}(p)$          &Log-Singularity (transverse)        & Factor \\[1mm]
\hline
0 & $-\frac{1}{16} p$                               & 0                                                 &       0                                                                          & 1   \\[1mm]
1 & $\frac{1}{N}\frac{13}{144\pi^2} p $  & $ \frac{1}{N}\frac{1}{24\pi^2} p$  &$\frac{1}{N}\frac{1}{12\pi^2} p\log{\frac{\Lambda}{p}} $  &     2     \\[1mm]
2 & $\frac{1}{N}\frac{1}{24\pi^2}     p $  & $-\frac{1}{N}\frac{1}{12\pi^2} p$  &  $-\frac{1}{N}\frac{1}{6\pi^2}   p\log{\frac{\Lambda}{p}} $&     1    \\[1mm]
3 & $\frac{1}{N}\frac{1}{3\pi^2}p$          & $-\frac{1}{N} \frac{2}{3\pi^2}p$    & $-\frac{1}{N}\frac{4}{3\pi^2}p\log{\frac{\Lambda}{p}}  $ &     2    \\[1mm]
4 & $-\frac{1}{N}0.110 p		      $         & $-\frac{1}{N} \frac{2}{3\pi^2}p$     &  0 &    1     \\[1mm]
5 & $\frac{1}{N} \frac{13}{18\pi^2}p$     & $\frac{1}{N} \frac{1}{\pi^2}p$        &  $\frac{1}{N} \frac{2}{3\pi^2}p\log{\frac{\Lambda}{p}}  $&   4     \\[1mm]
6 & $-\frac{1}{N} \frac{1}{4\pi^2}p$        & $-\frac{1}{N} \frac{1}{2\pi^2}p$     & 0 &   4     \\[1mm]
7 & $\frac{1}{N}\frac{1}{4\pi^2}p$          & $\frac{1}{N}\frac{1}{2\pi^2}p$       & 0 &   4      \\[1mm]
8 & $\frac{1}{N}\frac{5}{18\pi^2}p$        & $\frac{1}{N}\frac{1}{\pi^2}p$         & $\frac{1}{N}\frac{4}{3\pi^2} p\log{\frac{\Lambda}{p}}  $ &   2      \\[1mm]
9 & $-\frac{1}{N}\frac{13}{18\pi^2}p$     & $-\frac{1}{N} \frac{1}{\pi^2}p$       &$-\frac{1}{N}\frac{2}{3\pi^2} p\log{\frac{\Lambda}{p}} $   &   4     \\[1mm]
\hline
\end{tabular}
\caption{Evaluated contributions to the gauge-field self-energy at the critical point of the $CP^{N-1}$ model. 
Longitudinal components and logarithmic singularities both sum to 0.}
\label{tab:aa_crit}
\end{table}
Summing up the other contributions from Tab.~\ref{tab:aa_crit}, we find the renormalized gauge propagator 
\begin{align}
D_{\mu\nu}(p)=\frac{1}{\left(1/16 - \frac{0.036}{N}\right) p } \left( \delta_{\mu\nu} - \frac{\bp_\mu \bp_\nu}{p^2}\right)
\equiv \frac{C_A}{p }  \left( \delta_{\mu\nu} - \frac{\bp_\mu \bp_\nu}{p^2}\right)
\end{align}
with 
\begin{align}
C_A^{-1} = \frac{1}{16} - \frac{0.036}{N} = \frac{1}{16}\left(1 - \frac{0.578}{N}\right)\;.
\label{eq:C_A}
\end{align}
Utilizing the relation to the topological current mentioned above Eq.~(\ref{eq:C_A_intro}) 
and expanding to $1/N$ yields the value quoted in Eq.~(\ref{eq:C_A_intro}) (the difference in a global factor of $N$ 
is just a change of normalization mentioned above Eq.~(\ref{Z})). We are not aware of any previous computations of this number and it would be desirable to compare these results with other approaches.

\section{Extension of CP$^{N-1}$ model into symmetry-broken phase}
\label{sec:ssb}

In this section, we extend our analysis of the CP$^{N-1}$ model to the ``magnetic'' phase with spontaneously broken 
$S\!U(N)$ flavor symmetry 
(henceforth referred to as Goldstone phase). Our main motivation here is to lay the groundwork for our recently reported 
dynamics of the vector boson close to a deconfined quantum critical point \cite{huh13_short}. 

To derive an effective action for the Goldstone phase, we first choose 
the condensate $\sigma_0$ to be along the flavor index $i=1$ direction without loss of generality. It is convenient to use a radial 
coordinate system for the first flavor component so that
\begin{align}
z(\bx)=\left( \sigma(\bx)e^{i\omega(\bx)},
\pi_1(\bx),
\pi_2(\bx),
\dotsc,
\pi_{N-1}(\bx) \right)\;,
\label{eq:zsub}
\end{align}
where the $\pi_i$-fields are complex-valued and $\sigma(\bx)$ and $\omega(\bx)$ are real-valued. As a consequence of this 
coordinate transformation, the measure of the functional integral for the first flavor component at each point $\bx$ picks 
up a Jacobian \cite{munster},
\begin{align}
\mathcal{D}\{z^\ast(\bx),z(\bx)\} = \sigma(\bx) \mathcal{D}\{\sigma(\bx),\omega(\bx)\}\;.
\label{eq:jacobian}
\end{align}

In unitary gauge, 
the (redundant) local gauge transformation function is chosen as the phase variable of the first flavor $\omega(\bx)$ 
\cite{appelquist73,munster}. Then, as usual, the Goldstone boson of the first flavor is ``eaten up'' and the 
action does not depend on $\omega(\bx)$. Furthermore, we shift 
\begin{align}
\sigma \rightarrow \sqrt{N} \sigma_0 + \sigma\;,
\end{align}
and let the $\sigma$ field be the amplitude fluctuations around the condensate, with the condition
\begin{align}
\sigma_0^2 = \frac{1}{g} - \int_\bp \frac{1}{p^2}\;.
\end{align}
It is crucial to perform the shift in $\sigma$ also for the Jacobian, and re-exponentiate
it as a propagator $\langle \bar{c} c \rangle$ of fermionic ghost fields $\bar{c}$, $c$. The prefactor for the ghost Lagrangian is chosen 
such that the masses of the Goldstone bosons, $\Sigma_{\bar{\pi \pi}}(p=0)$, stay identically zero, as we will show below. 

The large-$N$ expansion can be performed by first integrating the $\pi$, $\bar{\pi}$ and expanding the still dynamical determinant 
to quadratic order in the fields $A_\mu$, $\sigma$ and $\lambda$. 
Executing all the before mentioned steps, we obtain the partition function
$Z = \int \mathcal{D} \sigma \mathcal{D} \lambda \mathcal{D}  A_\mu \exp \left( - \mathcal{S}_0 - \mathcal{S}_1 - \mathcal{S}_2 
-\mathcal{S}_3-\mathcal{S}_{\text{gh}} \right)$, with the action
\begin{align}
\mathcal{S}_0 &=
 \int_\bp  \left[ p^2 \sigma^2 + 2 i \sigma_0 \sigma \lambda + \frac{1}{2} \Pi(p,0) \lambda^2 + \sigma_0^2 A_\mu^2 
 + \frac{\Pi_A (p)}{2} \left( \delta_{\mu \nu} - \frac{p_\mu p_\nu}{p^2} \right) A_\mu A_\nu
 \right]\nn
\mathcal{S}_1 &= -\frac{i}{\sqrt{N}} \left( \frac{1}{g} - \sigma_0^2 \right) \int_\bx \lambda - \frac{1}{2N} \int_\bp \Pi (p,0) \lambda^2 
+ \frac{i}{\sqrt{N}} \int_\bx
  \lambda\, \sigma^2 \nn
&~~~~~+ \frac{2\sigma_0}{\sqrt{N}} \int_\bx A_\mu^2 \sigma + \frac{1}{N} \int_\bx A_\mu^2 \sigma^2  - \frac{1}{2N} \int_\bp
\Pi_A (p) \left( \delta_{\mu \nu} - \frac{p_\mu p_\nu}{p^2} \right) A_\mu A_\nu
\nonumber\\
\mathcal{S}_2 &= -\frac{i (N-1)}{3N^{3/2}} \int_{\bp_1,\bp_2,\bp_3} K_3 (\bp_1,\bp_2, \bp_3) \lambda(\bp_1) \lambda(\bp_2) \lambda (\bp_3)
\nn
&~~~~-  \frac{(N-1)}{12N^2} \int_{\bp_1,\bp_2,\bp_3,\bp_4} K_4 (\bp_1,\bp_2, \bp_3,\bp_4) \lambda(\bp_1) \lambda(\bp_2) \lambda (\bp_3) \lambda(\bp_4) 
\nonumber \\
&~~~~- \frac{i (N-1)}{3N^{3/2}} \int_{\bp_1,\bp_2,\bp_3}K_{A_\mu A_\nu \lambda} (\bp_1,\bp_2, \bp_3) A_\mu (\bp_1) A_\nu (\bp_2) \lambda (\bp_3) \nn
\mathcal{S}_3 &= \frac{N-1}{N^2} \int_{\bp_1,\bp_2,\bp_3,\bp_4} K_{\mu\nu\lambda\rho} (\bp_1,\bp_2, \bp_3,\bp_4) 
A_\mu (\bp_1) A_\nu (\bp_2) A_\lambda (\bp_3) A_\rho (\bp_4)
\nonumber\\
\mathcal{S}_{\text{gh}}&= \frac{\sigma_0}{\sqrt{N}}\int_\bx \bar{c} \left( \sqrt{N} \sigma_0+\sigma \right) c \;.
  \label{zhiggs} 
\end{align}

\begin{figure}[]
\includegraphics[width=50mm]{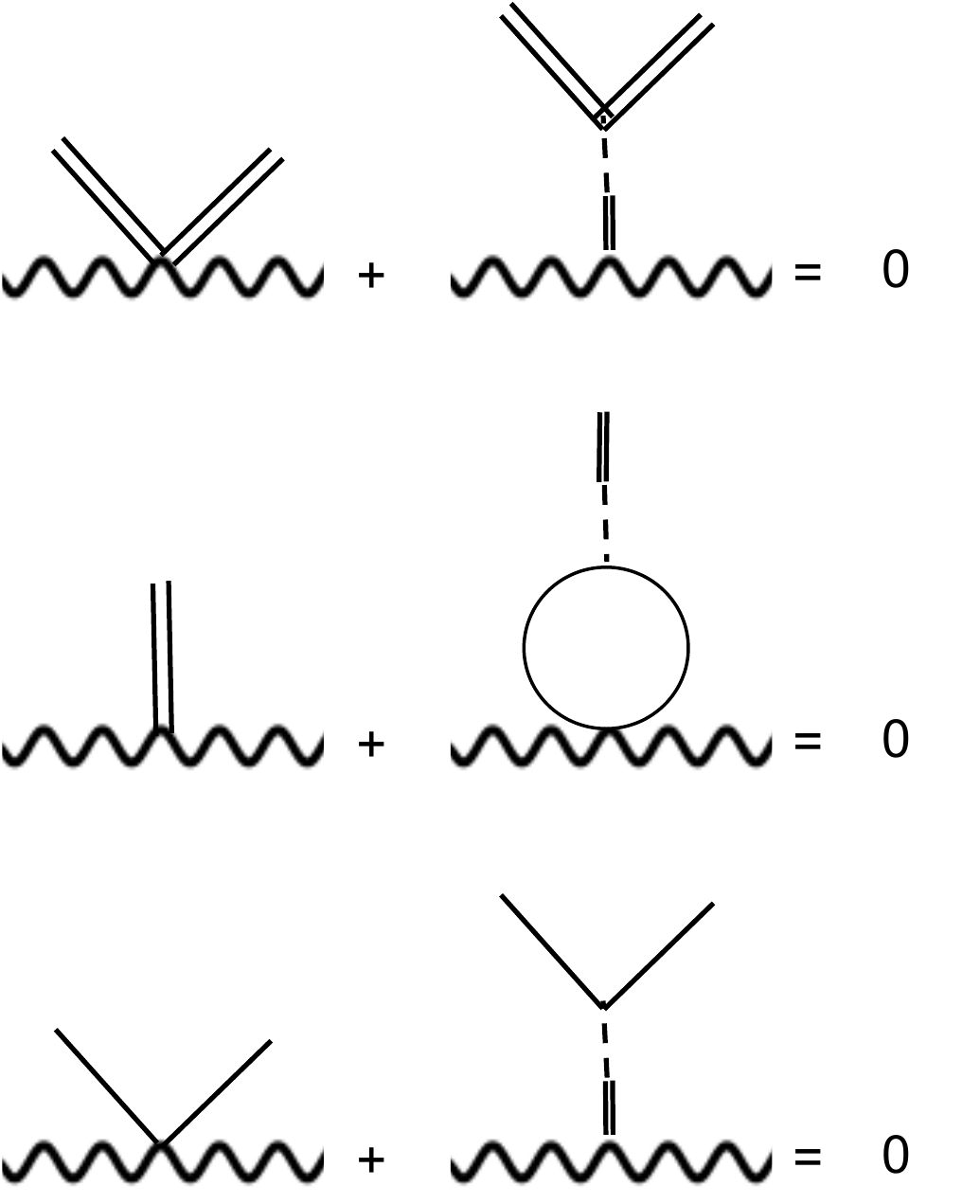}\\[5mm]
\includegraphics[width=120mm]{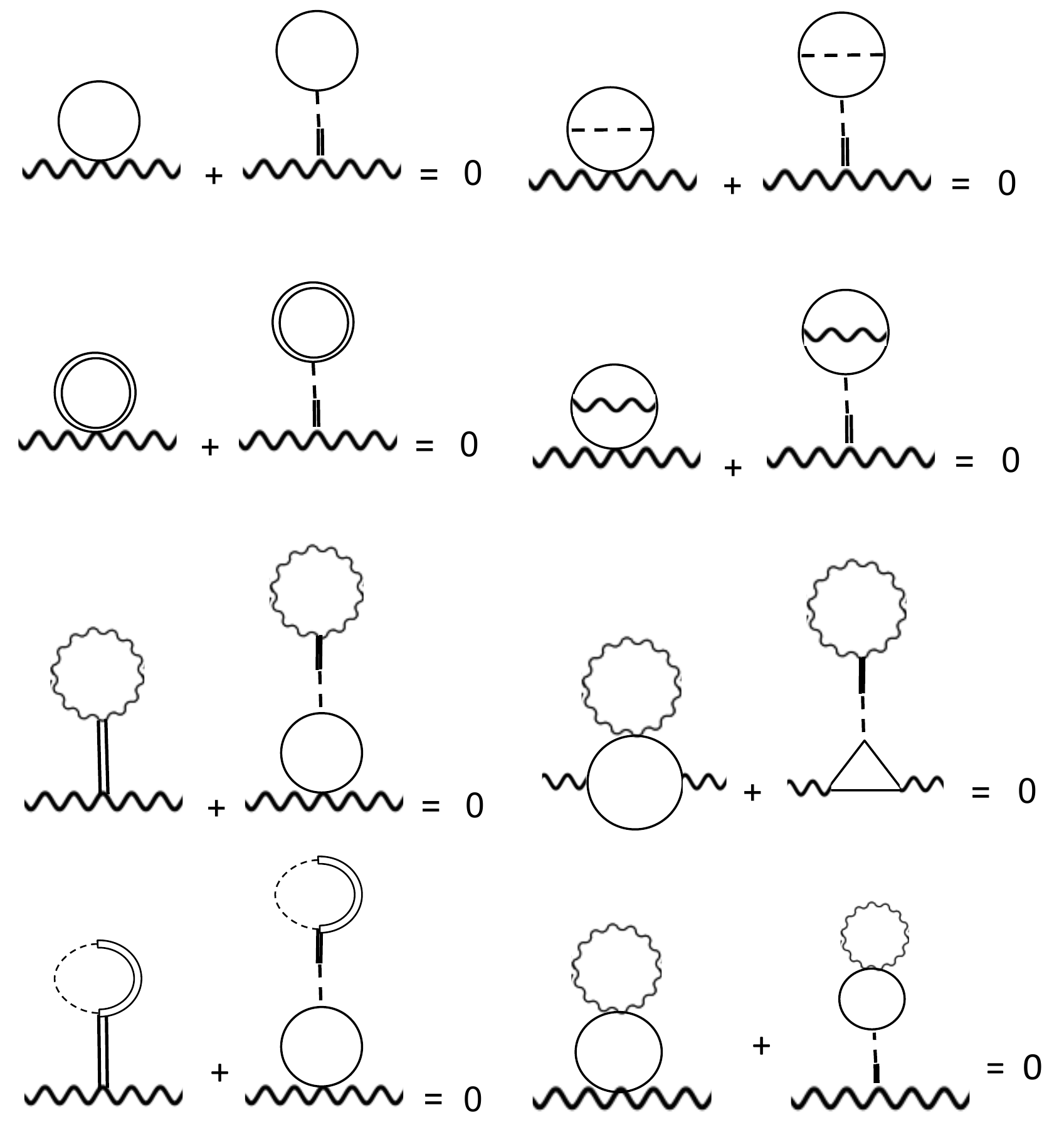}
\caption{Mutually canceling vertices (top) and loop contractions to order $1/N$ in the Goldstone phase.
After this cancellation, the only diagram that contributes to the gauge propagator is the polarization bubble, 
Fig.~\ref{fig:diags_gauge_crit}~(0), that is accounted for in Eq.~(\ref{eq:props2}).}
\label{fig:tadpoles}
\end{figure}

The various interaction vertices $K_{...}$ among the Lagrange multipliers $\lambda$ and gauge fields $A_\mu$, are generated 
by (closed) Goldstone boson loop diagrams. 
Using the abbreviations $\Pi (p,0) = \frac{1}{8p}$ and $\Pi_A (p) = \frac{p}{16}$,
the bare $N \rightarrow \infty$ Green's functions of the theory are
\bea
G_{\pi\bar{\pi}}^0(p) &=& \frac{1}{p^2} \nn
G_{\sigma\sigma}^0 (p) &=& \frac{\Pi (p,0)/2}{p^2 \Pi (p,0) + 2 \sigma_0^2}= \frac{1}{2p (p + 16  \sigma_0^2)} \nn 
G_{\lambda\lambda}^0 (p) &=& \frac{p^2}{p^2 \Pi (p,0) + 2 \sigma_0^2} = \frac{8p^2}{p + 16 \sigma_0^2} \nn 
G_{\sigma\lambda}^0 (p) &=& \frac{- i\sigma_0}{p^2 \Pi (p,0) + 2 \sigma_0^2}  = \frac{-8 i \sigma_0 }{p + 16 \sigma_0^2} \nn
D_{\mu\nu}^0 (p) &=& \frac{1}{2\sigma_0^2} \frac{p_\mu p_\nu}{p^2} + \frac{1}{(\Pi_A (p) + 2 \sigma_0^2)} 
\left( \delta_{\mu \nu} - \frac{p_\mu p_\nu}{p^2} \right)
= \frac{16}{(p + 32 \sigma_0^2)} \left( \delta_{\mu\nu} 
+ \frac{p_\mu p_\nu}{32 p \sigma_0^2} \right)\nn
G_{c\bar{c}}^0(p) &=& \frac{1}{\sigma_0^2}\;.
\label{eq:props2}
\eea
\begin{figure}[h]
\includegraphics[width=80mm]{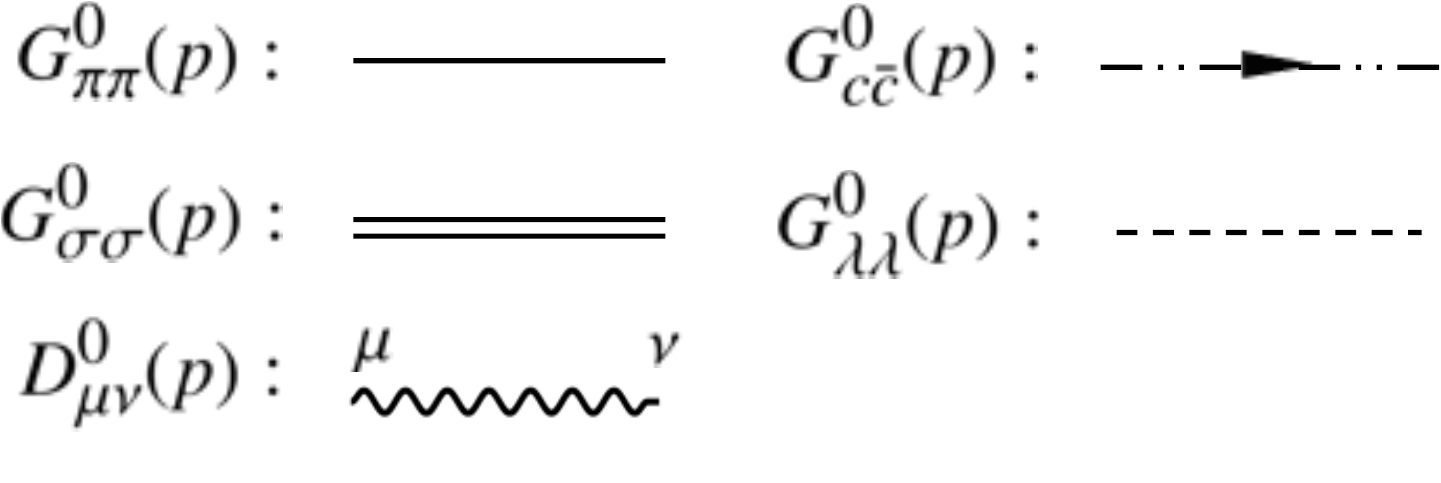}\\[-5mm]
\includegraphics[width=95mm]{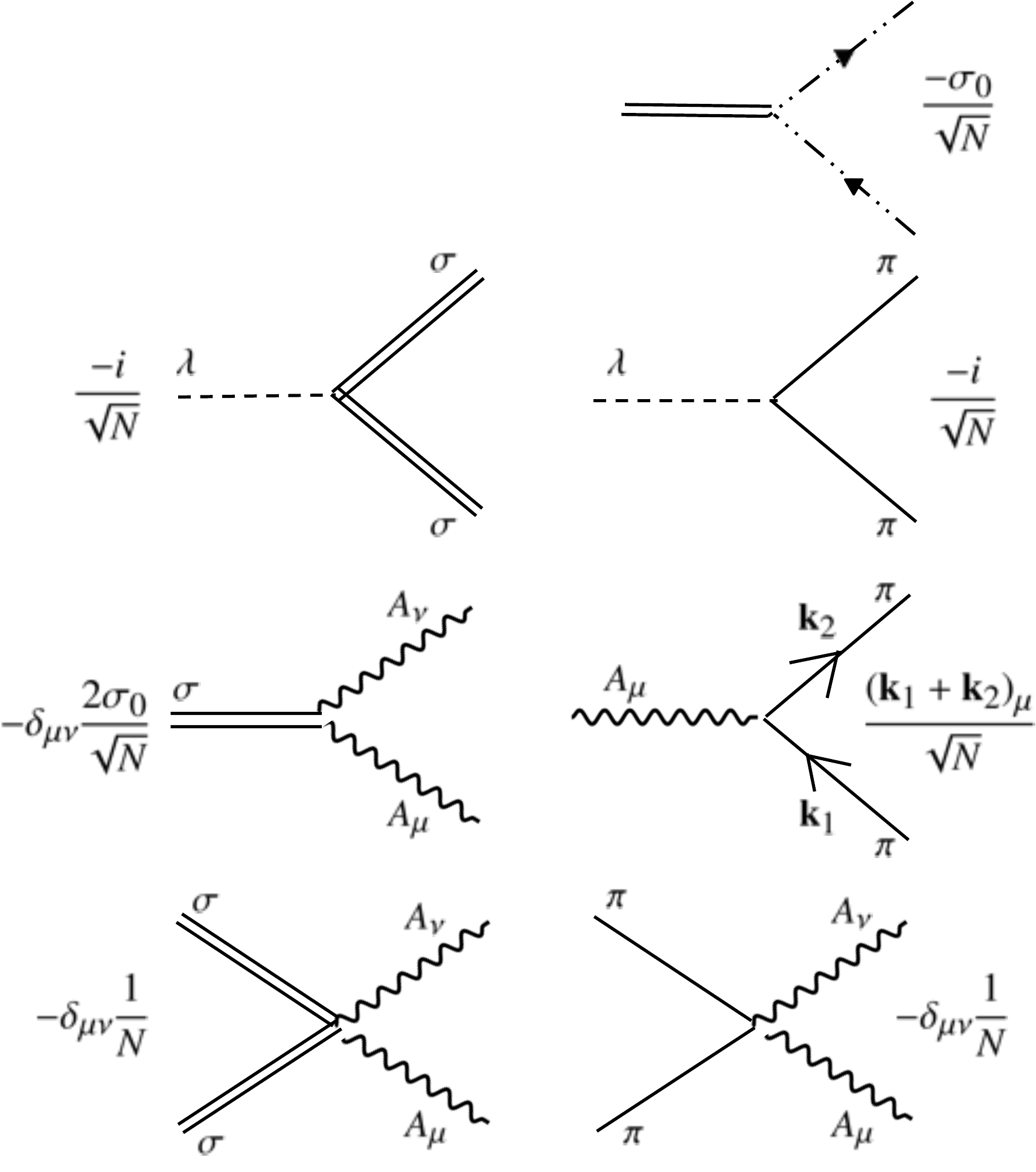}
\caption{Propagators and vertices of the $CP^{N-1}$ model in the symmetry broken phase.}
\label{fig:props}
\end{figure}
From the Feynman rules in Fig.~\ref{fig:props}, it can be seen that certain vertices cancel each other. These are shown at the top of 
Fig.~\ref{fig:tadpoles}. This figure also shows the mutually canceling diagrams renormalizing the gauge propagator, 
including a pair that each is of order unity ({\it i.e.}\ that would contribute even at $N\rightarrow \infty$). 
Due to this cancellation of the $N \rightarrow \infty$ diagrams, the only 
remaining large $N$ diagram of the gauge field propagator is the polarization bubble (0) shown in Fig.~\ref{fig:diags_gauge_crit}.

\subsection{Fulfillment of Goldstone's theorem to $1/N$}
\begin{figure}[t]
\includegraphics[width=75mm]{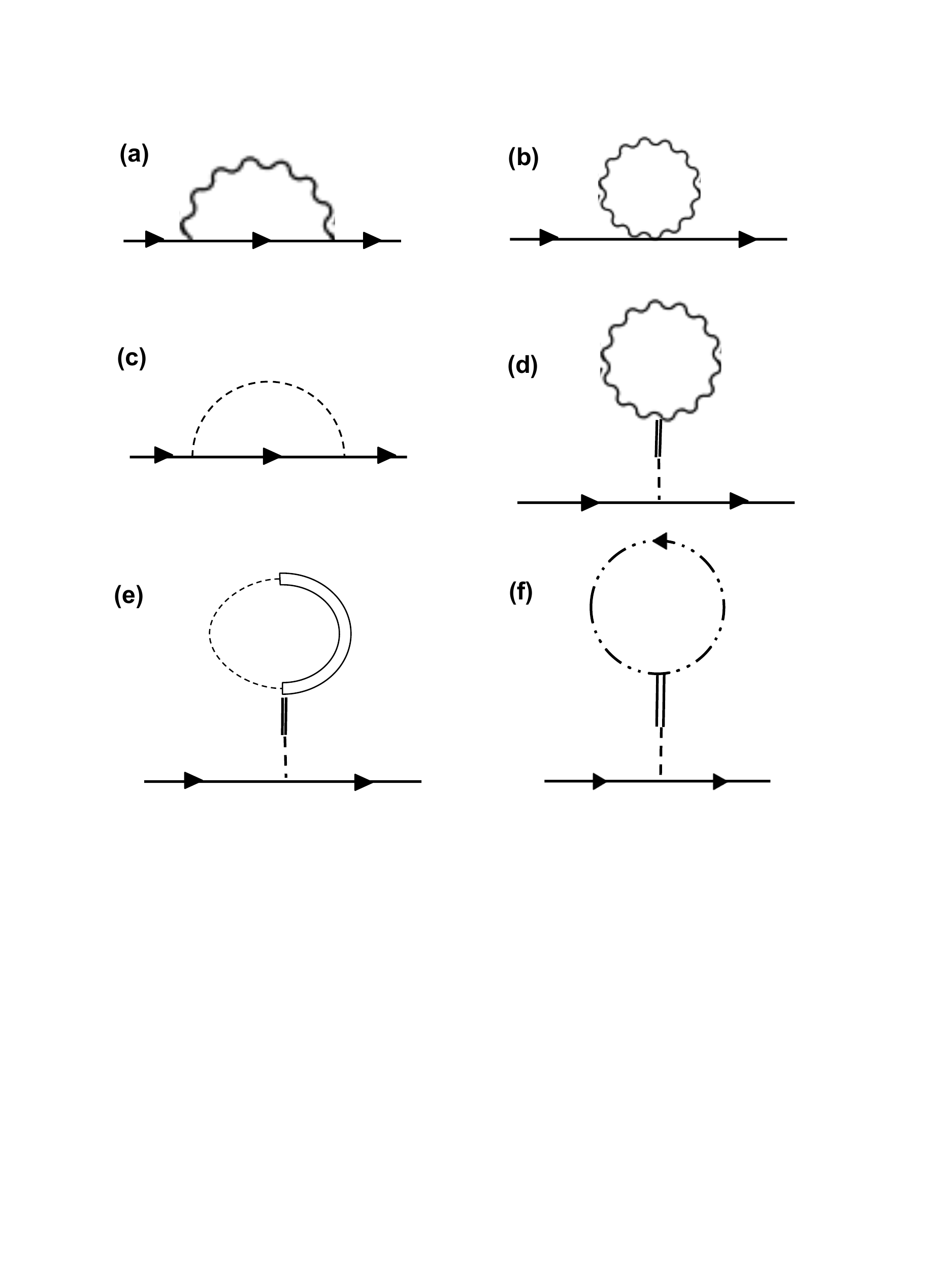}\\[-5mm]
\caption{Self energy contributions to the $\langle \bar{\pi} \pi \rangle$ Goldstone propagator at $1/N$ in the symmetry-broken phase.}
\label{fig:orderpi}
\end{figure}

In this section, we show that the contractions appearing at order $1/N$ in the large-N expansion do not generate 
unphysical masses for the Goldstone bosons. Within this approach, the mass of the Goldstone fields remains identically 
zero without invoking additional Ward identities. The diagrams that renormalize the propagators of the Goldstone bosons are in Fig. \ref{fig:orderpi}.
The following cancellations occur:
$2\Sigma_{\bar{\pi}\pi} ^{(b)}(p)+2\Sigma_{\bar{\pi}\pi}^{(d)}(p)=0$ can be seen using the third line in Fig. \ref{fig:tadpoles}. 
$\Sigma_{\bar{\pi}\pi}^{(c)}(0)+2\Sigma_{\bar{\pi}\pi}^{(e)}(0)=0$, and $\Sigma_{\bar{\pi}\pi}^{(a)}(0)+\Sigma_{\bar{\pi}\pi}^{(f)}(0)=0$ can be seen below. These can be seen explicitly from the expressions
\bea
\Sigma_{\bar{\pi}\pi}^{(a)}(0)&=&\frac{1}{N}\int_{ \bq} \frac{q_\mu q_\nu}{q^2}\frac{16}{q+32\sigma_0^2} 
\left( \delta_{\mu\nu}+\frac{q_\mu q_\nu}{32 q \sigma_0^2}\right)
= \frac{1}{N}\int_{ \bq} \frac{1}{2\sigma_0^2} \nn
\Sigma_{\bar{\pi}\pi}^{(c)}(0) &=& \left( -\frac{i}{\sqrt{N}}\right)^2 \int_\bq \frac{8q^2}{q+16\sigma_0^2}\frac{1}{q^2} \nn
\Sigma_{\bar{\pi}\pi}^{(e)}(0) &=& \left( -\frac{i}{\sqrt{N}}\right)^2 \int_\bq \frac{-8i\sigma_0}{16\sigma_0^2} \frac{-8i\sigma_0}{q+16\sigma_0^2} \nn
\Sigma_{\bar{\pi}\pi}^{(f)}(0) &=&- \frac{1}{N}\int_{ \bq} \frac{1}{2\sigma_0^2}\;.
\eea
With this, we arrive at the result
%
$\Sigma_{\bar{\pi}\pi}(0)=\sum_{i=a}^{f} a_i \Sigma_{\bar{\pi}\pi}^{(i)}(0)=0$\;
%
to conclude that the Goldstone mass remains 0 as it should. 
The importance of ghost field becomes obvious here (diagram (f) in Fig.~\ref{fig:orderpi}): without the 
ghosts the Goldstone boson would pick up an unphysical mass.

\subsection{Correlation length exponent $\nu$ and cancellation of singularities to $1/N$}

\begin{figure}[!ht]
\includegraphics[width=40mm]{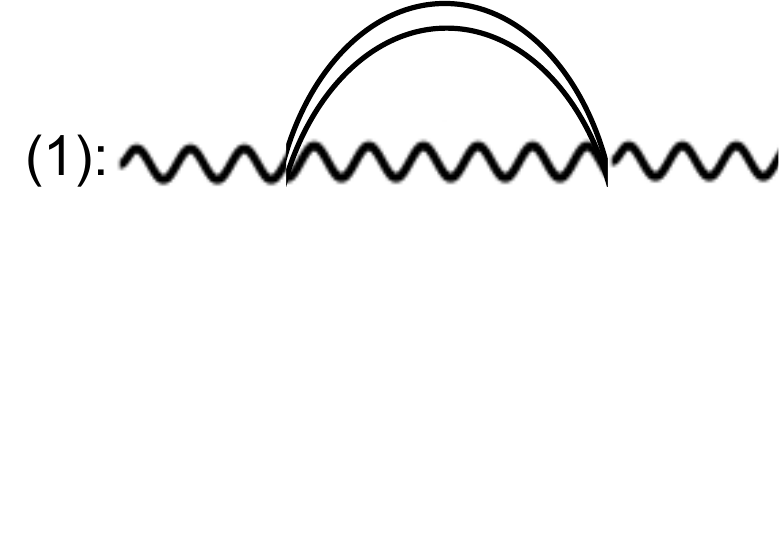}\\[-18mm]
\includegraphics[width=100mm]{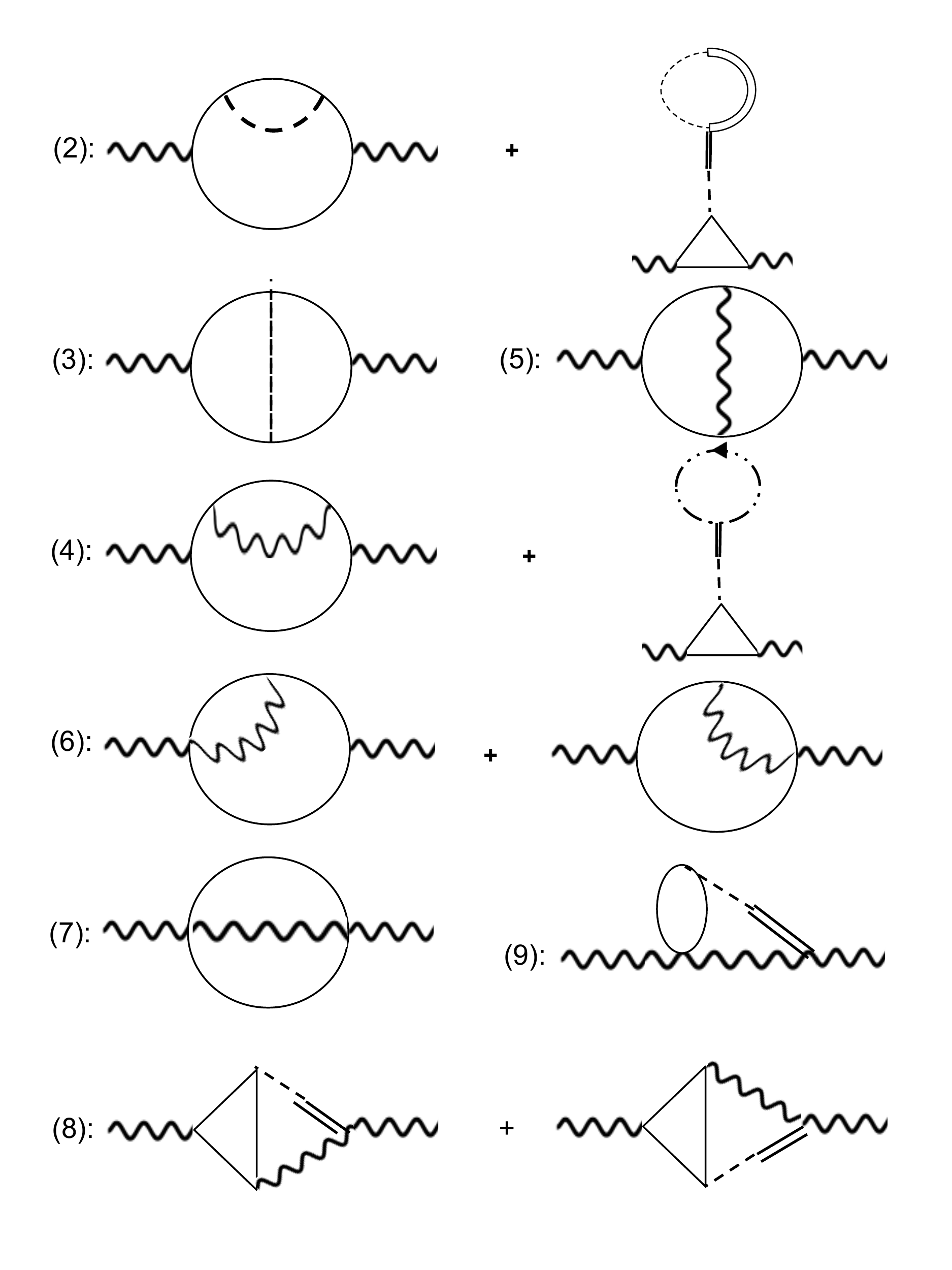}
\includegraphics[width=100mm]{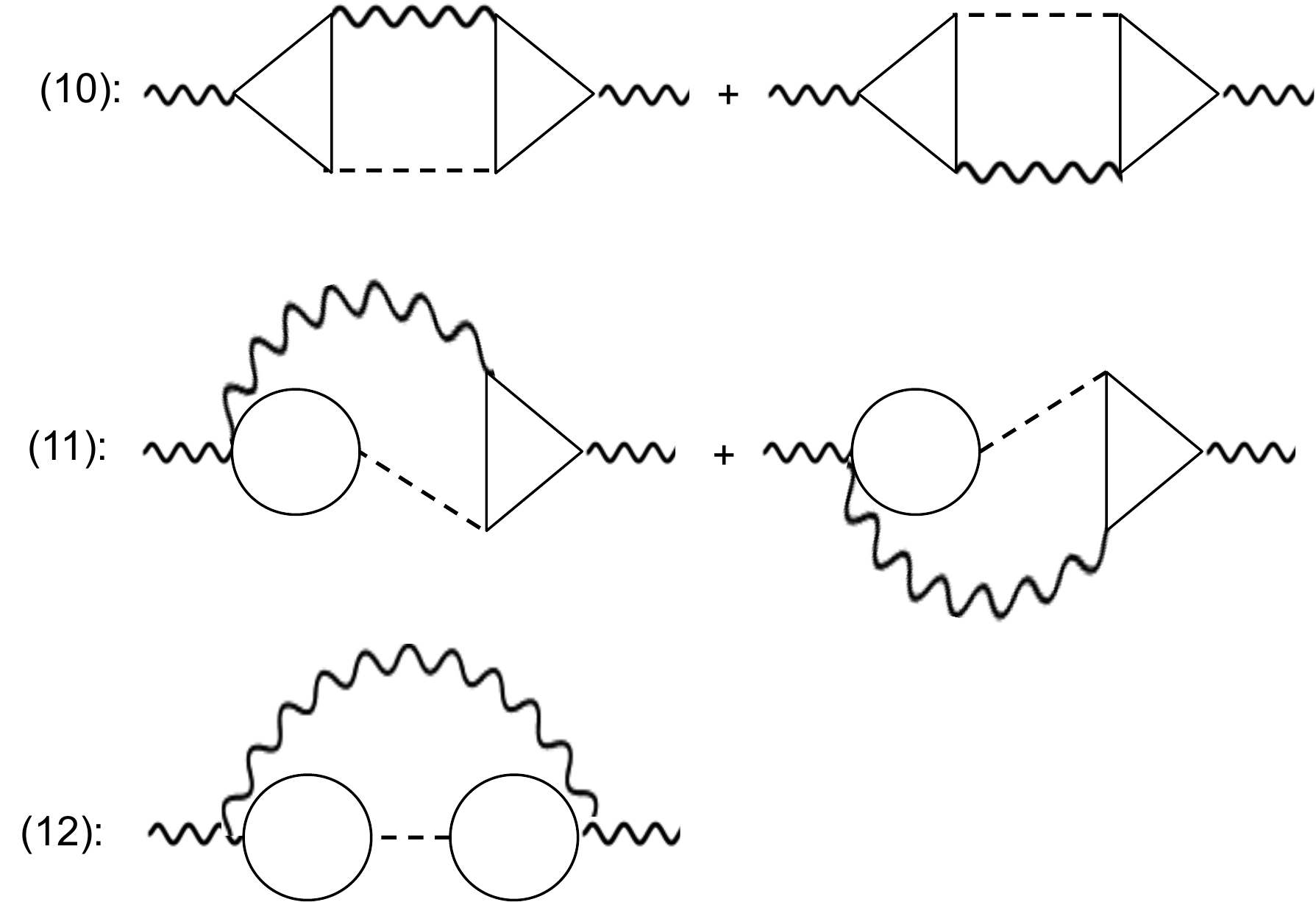}
\caption{1-, 2-, and 3-loop diagrams for the gauge propagator to order $1/N$ in the symmetry-broken phase.}
\label{fig:1loop}
\end{figure}

In this subsection, we compute the correlation length exponent $\nu$ from the symmetry-broken phase by resumming 
the log-singularities in the Higgs mass of the gauge propagator. We further show that all momentum dependent singularities from different diagrams mutually cancel to $1/N$. 
We write the self energy corrections to order $1/N$ shown in Fig.~\ref{fig:1loop} as 
\beq
\Sigma_{\mu\nu} (p) = \sum_{i=1}^{12} a_i \Sigma^{(i)}_{\mu\nu} (p)\;.
\eeq

The contributions from each diagram are
\begin{alignat}{2}
\Sigma^{(1)}_{\mu\nu} (p) &= \frac{32 \sigma_0^2}{N} \int_\bq \left( \delta_{\mu\nu} 
  + \frac{q_\mu q_\nu}{32 q \sigma_0^2} \right)\frac{1}{(q + 32 \sigma_0^2)|\bp + \bq| (|\bp + \bq| + 16  \sigma_0^2)} 
  & a_1=4\nn
\Sigma^{(2)}_{\mu\nu} (p) &= -\frac{8}{N} \int_\bq \int_\bk 
   \frac{(2\bk+\bp)_\mu (2\bk+\bp)_\nu q^2 }{ k^4 (\bp+\bk)^2  (q+16   \sigma_0^2 )} \left( \frac{1}{(\bk+\bq)^2} - \frac{1}{q^2}\right) 
  & a_2=2\nn
\Sigma^{(3)}_{\mu\nu} (p) &= -\frac{8}{N} \int_\bq \int_\bk 
   \frac{(2\bk+\bp)_\mu ( 2(\bk+\bq)+\bp)_\nu q^2 }{ k^2 (\bp+\bk)^2 (\bk+\bq)^2 (\bp+\bk+\bq)^2 (q+16   \sigma_0^2 )} 
   & a_3=1\nn
\Sigma^{(4)}_{\mu\nu} (p) &= \frac{16}{N} \int_\bq \int_\bk 
   \frac{(2 \bk+\bp)_\mu (2\bk+\bp)_\nu  }{ k^4 (\bp+\bk)^2(q+32 \sigma_0^2 )}
   \left( \frac{ (2\bk+\bq)_\lambda (2\bk+\bq)_\rho }{  (\bk+\bq)^2 }  -  \frac{ q_\lambda q_\rho }{q^2} \right) 
   \left( \delta_{\lambda\rho} + \frac{q_\lambda q_\rho}{32 q \sigma_0^2} \right) 
   & a_4=2\nn
\Sigma^{(5)}_{\mu\nu} (p) &= \frac{16}{N} \int_\bq \int_\bk 
   \frac{(2 \bk+\bp)_\mu (2 (\bk + \bq) +\bp)_\nu (2 (\bk + \bp)+\bq)_\lambda (2\bk+\bq)_\rho }{ k^2 (\bp+\bk)^2 (\bk+\bp+\bq)^2 (\bk+\bq)^2 (q+32 \sigma_0^2 )}
   \left( \delta_{\lambda\rho} + \frac{q_\lambda q_\rho}{32 q \sigma_0^2} \right) 
  & ~~a_5=1\nn
\Sigma^{(6)}_{\mu\nu} (p) &= -\frac{16}{N} \int_\bq \int_\bk 
  \frac{(2\bk+\bq)_\rho (2\bk+\bp)_\nu }{ k^2 (\bk+\bq)^2 (\bk+\bp)^2 (q+ 32 \sigma_0^2 )} 
  \left( \delta_{\lambda\rho} + \frac{q_\lambda q_\rho}{32 q \sigma_0^2} \right) \delta_{\mu\lambda} 
  + (\mu \leftrightarrow \nu) 
  & a_6 =4 \nn
\Sigma^{(7)}_{\mu\nu} (p) &= \frac{16}{N} \int_\bq \int_\bk 
  \frac{1}{ (\bk+\bq)^2 ( \bk + \bp )^2 ( q + 32 \sigma_0^2) }  
  \left( \delta_{\lambda\rho} + \frac{q_\lambda q_\rho}{32 q \sigma_0^2} \right) \delta_{\mu\lambda} \delta_{\nu\rho}
  & a_7=4 \nn
\Sigma^{(8)}_{\mu\nu} (p) &= \frac{256}{N} \int_\bq \int_\bk 
 \left( \frac{(2\bk+\bp)_\mu (2\bk+2\bp+\bq)_\lambda}{k^2(\bk+\bp)^2(\bk+\bp+\bq)^2} 
  + \frac{(2\bk+\bp)_\mu (2\bk-\bq)_\lambda}  {k^2(\bk+\bp)^2(\bk-\bq)^2} \right)\times 
  \nn
  &
 \hspace{25mm} \frac{\sigma_0^2 \delta_{\rho\nu} }{(|\bp+\bq|+16\sigma_0^2)(q+32\sigma_0^2)}  
  \left( \delta_{\lambda\rho} + \frac{q_\lambda q_\rho}{32 q \sigma_0^2} \right)
  & a_8=2 \nn
\Sigma^{(9)}_{\mu\nu} (p) &= -\frac{256\sigma_0^2}{N} \int_\bq \int _\bk
 \frac{1}{k^2(\bk+\bp+\bq)^2}\frac{1}{(|\bp+\bq|+16\sigma_0^2)(q+32\sigma_0^2)}
  \left( \delta_{\mu\nu}+ \frac{q_\mu q_\nu}{32 q \sigma_0^2} \right)
  & a_9=4 \nn
\Sigma^{(10)}_{\mu\nu} (p) &= -\frac{128}{N}  \int_{\bq,\bk,\bl} \Bigg[ 
  \frac{(2\bk+\bp)_\mu (2\bk+\bq)_\lambda }{ k^2 (\bk+\bp)^2 (\bk+\bq)^2} 
  \frac{(2\bl+\bq)_\rho (2\bl+\bp)_\nu }{ l^2 (\bl+\bp)^2 (\bl+\bq)^2} 
  \frac{(\bp-\bq)^2}{(|\bp-\bq|+16\sigma_0^2)( q + 32 \sigma_0^2) }  
  \left( \delta_{\lambda\rho} + \frac{q_\lambda q_\rho}{32 q \sigma_0^2} \right)\nn
  &
  \hspace{25mm}+\frac{(2\bk+\bp)_\mu (2\bk+\bp+\bq)_\lambda }{ k^2 (\bk+\bp)^2 (\bk+\bq)^2} 
  \frac{(2\bl+\bp+\bq)_\rho (2\bl+\bp)_\nu }{ l^2 (\bl+\bp)^2 (\bl+\bq)^2} 
  \frac{q^2}{(q+16\sigma_0^2)( |\bp-\bq|+ 32 \sigma_0^2) }  
  \times
  \nn
  &
 \hspace{25mm} \left( \delta_{\lambda\rho} + \frac{(\bp-\bq)_\lambda (\bp-\bq)_\rho}{32 |\bp-\bq| \sigma_0^2} \right)
  \Bigg]
   & a_{10}=2 \nn
\Sigma^{(11)}_{\mu\nu} (p) &= \frac{128}{N} \int_{\bq,\bk,\bl}
 \frac{ \delta_{\mu\lambda}}{k^2(\bk+\bp+\bq)^2} \left( \frac{(2\bl-\bq)_\rho(2\bl+\bp)_\nu}{l^2(\bl-\bq)^2(\bl+\bp)^2} 
  +  \frac{(2\bl+2\bp+\bq)_\rho(2\bl+\bp)_\nu}{l^2(\bl+\bp)^2(\bl+\bp+\bq)^2}\right)
  \nn
  &
 \hspace{22mm} \frac{(\bp+\bq)^2}{(q+32\sigma_0^2)(|\bp+\bq|+16\sigma_0^2)}
  \left( \delta_{\lambda\rho} + \frac{q_\lambda q_\rho}{32 q \sigma_0^2} \right)
  & a_{11}=4 \nn
\Sigma^{(12)}_{\mu\nu} (p) &= -\frac{128}{N} \int_{\bq,\bk,\bl}  
  \frac{1}{k^2(\bk+\bp+\bq)^2}\frac{1}{l^2(\bl+\bp+\bq)^2}\frac{(\bp+\bq)^2}{(|\bp+\bq|+16\sigma_0^2)(q+32\sigma_0^2)}
  \left( \delta_{\mu\nu}+ \frac{q_\mu q_\nu}{32 q \sigma_0^2} \right)
  & a_{12}=4 .\nn
  \label{eq:12}
\end{alignat}

\begin{table} [h]
\centering
\begin{tabular}{cccc }
\text{Diagram} &~ $N \Sigma_T^{UV}(p)$ ~~~& $N \Sigma_L^{UV}(p)$ & Factor \\[1mm]
\hline
1 & $\frac{\Lambda}{6 \pi^2} + \frac{8 \sigma_0^2}{\pi^2} \log{\frac{\Lambda}{32\sigma_0^2}} $  
   & $\frac{\Lambda}{6 \pi ^2} +  \frac{8 \sigma_0^2}{\pi ^2}\log{\frac{\Lambda}{32\sigma_0^2}} $  &  4     \\[1mm]
2 & $\frac{ p }{ 12 \pi^2 }\log{\frac{\Lambda}{32\sigma_0^2}}$  
   & $0$  &   2    \\[1mm]
3 & $-\frac{ p }{ 6 \pi^2 }\log{\frac{\Lambda}{32\sigma_0^2}}$          
   & $0$    &   1    \\[1mm]
4 & $\frac{8 \Lambda}{3\pi^2} + \frac{p\Lambda}{64\pi^2\sigma_0^2} + \left ( -\frac{p^2}{120 \pi ^2 \sigma _0^2}-\frac{4 p}{3 \pi ^2}-\frac{256 \sigma _0^2}{3\pi^2}\right)  \log{\frac{\Lambda}{32\sigma_0^2}}$         
   & $ \frac{8\Lambda}{3 \pi ^2}+\left ( \frac{p^2}{160 \pi ^2 \sigma_0^2 }-\frac{256\sigma_0^2}{3 \pi^2 } \right) \log{\frac{\Lambda}{32\sigma_0^2}}$     &  2     \\[1mm]
5 & $ \frac{4 \Lambda}{3\pi^2} - \frac{p \Lambda}{ 32 \pi^2\sigma_0^2} + \frac{\Lambda^2}{48 \pi^2 \sigma_0^2} + \left(\frac{p^2}{120 \pi^2 \sigma_0^2} - \frac{128 \sigma_0^2}{3\pi^2} \right)\log{\frac{\Lambda}{32\sigma_0^2}}$     
   & $ \frac{4\Lambda}{3 \pi ^2}+\frac{\Lambda^2} {48 \pi^2 \sigma_0^2 }+ \left ( \frac{p^2}{240 \pi ^2 \sigma_0^2 }-\frac{128 \sigma_0^2}{3 \pi^2 } \right) \log{\frac{\Lambda}{32\sigma_0^2}}$      &  1     \\[1mm]
6 & $ -\frac{4\Lambda}{3\pi^2} - \frac{\Lambda^2}{96\pi^2 \sigma_0^2} + \left( \frac{2p}{3\pi^2} + \frac{128 \sigma_0^2}{3\pi^2}+\frac{p^2}{240 \pi^2 \sigma_0^2} \right) \log{\frac{\Lambda}{32\sigma_0^2}}$        
  & $ -\frac{4\Lambda}{3 \pi ^2}-\frac{\Lambda^2} {96 \pi^2 \sigma_0^2 }+ \left (- \frac{p^2}{120 \pi ^2 \sigma_0^2 }+\frac{128 \sigma_0^2}{3 \pi^2 } \right) \log{\frac{\Lambda}{32\sigma_0^2}}$     &   4     \\[1mm]
7 & $ \frac{2\Lambda}{3\pi^2} + \frac{\Lambda^2}{192\pi^2 \sigma_0^2} + \left(- \frac{p^2}{480 \pi^2 \sigma_0^2}- \frac{64 \sigma_0^2}{3 \pi^2}   \right) \log{\frac{\Lambda}{32\sigma_0^2}}$     
  & $\frac{2\Lambda}{3 \pi ^2}+\frac{\Lambda^2} {192 \pi^2 \sigma_0^2 }+ \left ( \frac{p^2}{240 \pi ^2 \sigma_0^2 }-\frac{64 \sigma_0^2}{3 \pi^2 } \right) \log{\frac{\Lambda}{32\sigma_0^2}}$        &   4      \\[1mm]
8 & $\frac{\Lambda}{3\pi^2}+\frac{16\sigma_0^2}{\pi^2}\log{\frac{\Lambda}{32\sigma_0^2}}$        
   & $\frac{\Lambda}{3\pi^2} + \frac{16\sigma_0^2}{\pi^2}\log{\frac{\Lambda}{32\sigma_0^2}}$         &   2      \\[1mm]
9 & $-\frac{\Lambda}{6\pi^2}-\frac{8\sigma_0^2}{\pi^2}\log{\frac{\Lambda}{32\sigma_0^2}}$          
   & $-\frac{\Lambda}{6\pi^2} - \frac{8\sigma_0^2}{\pi^2}\log{\frac{\Lambda}{32\sigma_0^2}}$      &    4     \\[1mm]
10 & $-\frac{\Lambda}{\pi^2} - \frac{\Lambda^2}{96\pi^2\sigma_0^2}+ \left( \frac{4 p}{3\pi^2} + \frac{p^2}{240\pi^2\sigma_0^2} + \frac{176\sigma_0^2}{3\pi^2} \right)\log{\frac{\Lambda}{32\sigma_0^2}} $          
    & $-\frac{\Lambda}{\pi^2} - \frac{\Lambda^2}{96\pi^2\sigma_0^2} + \left( -\frac{p^2}{120\pi^2\sigma_0^2} + \frac{176\sigma_0^2}{3\pi^2} \right) \log{\frac{\Lambda}{32\sigma_0^2}} $      &    2     \\[1mm]
11 & $\frac{\Lambda}{\pi^2} + \frac{\Lambda^2}{96\pi^2\sigma_0^2}+ \left( -\frac{2 p}{3\pi^2} - \frac{p^2}{240\pi^2\sigma_0^2} - \frac{176\sigma_0^2}{3\pi^2} \right)\log{\frac{\Lambda}{32\sigma_0^2}}$          
    & $\frac{\Lambda}{\pi^2} + \frac{\Lambda^2}{96\pi^2\sigma_0^2} + \left( \frac{p^2}{120\pi^2 \sigma_0^2} - \frac{176\sigma_0^2}{3\pi^2} \right) \log{\frac{\Lambda}{32\sigma_0^2}} $      &    4     \\[1mm]
12 & $-\frac{\Lambda}{2\pi^2}-\frac{\Lambda^2}{192 \pi^2\sigma_0^2} + \left( \frac{p^2}{480\pi^2\sigma_0^2} + \frac{88\sigma_0^2}{3\pi^2} \right) \log{\frac{\Lambda}{32\sigma_0^2}}$          
     & $-\frac{\Lambda}{2\pi^2} - \frac{\Lambda^2}{192\pi^2\sigma_0^2} + \left( -\frac{p^2}{240\pi^2\sigma_0^2} + \frac{88\sigma_0^2}{3\pi^2} \right) \log{\frac{\Lambda}{32\sigma_0^2}} $      &    4     \\[2mm]
\hline
\end{tabular}
\caption{Evaluated contributions to the divergent parts of transverse and longitudinal gauge-field self-energies in the symmetry broken phase of the $CP^{N-1}$ model. All momentum dependent terms cancel and all that remains is Eq.~(\ref{eq:UVdiv}). }
\label{tab:nu}
\end{table}

We now extract the divergent terms in the Goldstone phase using Tensoria (cf. App.~\ref{app:tensoria}). Resumming the logarithmically divergent terms (Table~\ref{tab:nu}), we get 
\bea
\Sigma^{UV}_{\mu\nu} (p) = \frac{1}{N}\left( \frac{14 \Lambda}{3\pi^2} - \frac{96 \sigma_0^2}{ \pi^2}  \log{\frac{\Lambda}{32\sigma_0^2}}  \right ) \delta_{\mu\nu}\;.
\label{eq:UVdiv}
\eea
This comes in the renormalized gauge propagator as 
\bea
2\sigma_0^2 \delta_{\mu\nu} \left( 1+ \frac{1}{N}\frac{48}{\pi^2}\log{\frac{\Lambda}{32\sigma_0^2}}   \right) \;.
\eea
Thus the correlation length exponent is
\bea
\nu = 1-\frac{48}{N \pi^2}\;,
\eea
which is consistent with known $CP^{N-1}$ results \cite{halperin74,irkhin96}. We note that each individual diagram 
features other types of singular terms: of the form $p\log{\Lambda}$ and $p^2/\sigma_0^2 \log{\Lambda}$. 
It is only after summing all the diagrams that these cancel with each other.

An application of the results presented in this Section \ref{sec:ssb} is a computation of the excitation 
spectrum of the vector boson near the critical point, which has been reported recently \cite{huh13_short}.

\section{Conclusion}

In this paper, we computed response functions of conserved vector currents in the CP$^{N-1}$ model with a view towards 
applying these results to the physics of deconfined quantum criticality. Vector response functions are 
of interest for quantum critical transport, {\it i.e.}\ the dynamic response of quantum spin systems subject 
to magnetic fields, as well as to identify the presence of fractionalized gauge excitations at the critical point.
Our main objective was to provide a new set of quantitative predictions that allow numerical simulations, 
and ultimately experiments, of quantum spin systems to discriminate between a conventional O(3) versus deconfined 
CP$^{1}$ critical point.

We first computed universal amplitudes of the current-current correlator $\langle J_\mu J_\nu \rangle$ (magnetic transport) and the 
gauge propagator $\langle A_\mu A_\nu \rangle$ (related to the topological current) in the conformally invariant regime at the critical point. 
Going to order $1/N$ in a large-$N$ expansion, we clarified the diagrammatic structure in momentum space by demonstrating explicit 
cancellations of singularities that would otherwise have violated the (exact) constraints imposed by conformal symmetry in $2+1$ dimensions. To achieve this, we developed an algorithm to reliably evaluate tensor-valued momentum integrals by relating 
them to scalar integrals using Davydychev permutation relations.

We then extended our theory for the CP$^{N-1}$ model to the phase with spontaneously broken $S\!U(N)$ flavor symmetry 
thereby providing the groundwork to investigate the nature of vector boson excitations in this regime. 
The flavor condensate results in a `Higgsed', massive gauge boson and complicates the propagation and interaction 
channels for the spinons. The $1/N$-approach, in combination with fixed unitary gauge and fermionic ghost fields, 
was shown to be consistent with Ward identities/Goldstone's theorem and enabled us to access the critical behavior of the 
vector boson from the symmetry-broken side of the critical point.

Going forward, we hope that the diagrammatic structure made transparent in this paper becomes helpful also in 
some long-standing problems of the correlated electron community, such as capturing the effects of order parameter 
fluctuations in two-dimensional superconductors.

\section{Acknowledgments}

We acknowledge helpful discussions with Debanjan Chowdhury and Matthias Punk. 
This research was supported by the DFG under grant Str 1176/1-1, by the NSF under Grant DMR-1103860, 
by the John Templeton foundation, by the Center for Ultracold Atoms (CUA) and by the Multidisciplinary University Research Initiative (MURI). 
This research was also supported in part by Perimeter Institute for
Theoretical Physics; research at Perimeter Institute is supported by the
Government of Canada through Industry Canada and by the Province of
Ontario through the Ministry of Research and Innovation.

\appendix

\section{Tensoria algorithm for tensor-valued momentum integrals}
\label{app:tensoria}

In this appendix, we explain in detail how we evaluate the momentum integrals using Tensoria.
As an example, we step through the evaluation of the integral $J^{\ell m}_{\mu\nu}(p)^{(4)}$  in 
Eq.~(\ref{eq:current_gauge_on}),
\begin{align}
J_{\mu\nu} (p)^{(4)} &=
 \frac{16}{N} 
 \int_\bq \int_\bk 
   \frac{(2 \bk+\bp)_\mu (2 (\bk + \bq) +\bp)_\nu (2 (\bk + \bp)+\bq)_\lambda (2\bk+\bq)_\rho }{ k^2 (\bp+\bk)^2 (\bk+\bp+\bq)^2 (\bk+\bq)^2}
  \left(\frac{ \delta_{\lambda\rho} q^2 - q_\lambda q_\rho}{ q^3} \right)\;.
  \label{eq:example}
\end{align}
This is the flavor diagonal part after having performed the flavor trace. We first perform the integration over 
$\mathbf{k}$. All our self-energy corrections of the form Table \ref{tab:c_j} are symmetric in the indices $\mu$, $\nu$ and the 
resulting basis of $3\times3$ matrices can be spanned by projections onto $J (p)^{(4)} \equiv \delta_{\mu\nu}J_{\mu\nu} (p)^{(4)}$ and 
$ p_{\mu}p_{\nu}/|\mathbf{p}|^2 J_{\mu\nu} (p)^{(4)}$. We continue with $J (p)^{(4)}$ as an example.
Expanding out the numerator of  $J^{(4)}(p)$, we get
\begin{align}
&16 K^2 Q^2 k_\nu p_\lambda+16 K^2 Q^2 k_\rho
 p_\rho-8 K^2 P^2 k_\rho q_\rho-16
   K^2 k_\lambda k_\rho q_\lambda
   q_\rho-8 K^2 Q^2 k_\lambda q_\lambda
   \nonumber\\
   &+16 K^2 Q^2 k_\nu q_\nu+8 K^2 R^2
   k_\rho q_\rho-8 P^2 k_\nu k_\rho
   p_\nu q_\rho+8 P^2 Q^2 k_\rho p(\rho
   )
   \nonumber\\
   &-16 k_\lambda k_\nu k_\rho p_\nu
   q_\lambda q_\rho-8 Q^2 k_\lambda
   k_\nu p_\nu q_\lambda+16 Q^2 k_\nu
   k_\rho p_\rho q_\nu+8 R^2 k_\nu
   k_\rho p_\nu q_\rho
   \nonumber\\
 &
  +4 Q^4 k_\rho
   p_\rho+16 Q^2 k_\nu k_\rho p_\nu
   p_\rho-4 Q^2 R^2 k_\rho p_\rho-4 P^4
   k_\rho q_\rho-8 P^2 k_\lambda k(\rho
   ) q_\lambda q_\rho
   \nonumber\\
   &
   -8 P^2 k_\nu
   k_\rho q_\nu q_\rho-4 P^2 Q^2
   k_\lambda q_\lambda-2 P^2 Q^2 k_\rho
   q_\rho+6 P^2 R^2 k_\rho q_\rho-16
   k_\lambda k_\nu k_\rho q_\lambda
   q_\nu q_\rho
   \nonumber\\
   &-2 Q^4 k_\lambda
   q_\lambda-8 Q^2 k_\lambda k_\nu
   q_\lambda q_\nu-4 Q^2 k_\lambda
   k_\rho q_\lambda q_\rho+2 Q^2 R^2
   k_\lambda q_\lambda+2 Q^2 R^2 k_\rho
   q_\rho
   \nonumber\\
  & -2 R^4 k_\rho q_\rho+4 R^2
   k_\lambda k_\rho q_\lambda q_\rho
  +8 R^2 k_\nu k_\rho q_\nu q_\rho
   +16 K^4 Q^2+8 K^2 P^2 Q^2+4 K^2 Q^4-4 K^2
   Q^2 R^2
   \label{eq:numerator}
\end{align}
with the absolute values of momenta denoted by capital letters $Q=|\mathbf{q}|$, $P=|\mathbf{p}|$, and 
$K=|\mathbf{k}|$, $R=|\mathbf{p}-\mathbf{q}|$; below we will also use $S=|\mathbf{p}+\mathbf{q}|$.

The next step is to transform the denominator containing four propagators to a sum of terms 
containing only three propagators using the identity
\begin{align}
 \frac{1}{ k^2 (\bp+\bk)^2 (\bk+\bp+\bq)^2 (\bk+\bq)^2}
 =
 \frac{1}{2 \mathbf{p}\cdot \mathbf{q}} 
 \Bigg[&
 \frac{1}{\left(\bk + \bp\right)^2 \left(\bk + \bq\right)^2
 \left(\bk + \bp + \bq\right)^2}
 +
  \frac{1}{\bk^2 \left(\bk + \bp\right)^2
 \left(\bk + \bq \right)^2}
 \nonumber\\
 &-
 \frac{1}{\bk^2 \left(\bk + \bq \right)^2 
  \left(\bk +\bp+ \bq \right)^2
 }
 -
  \frac{1}{\bk^2   \left(\bk + \bp \right)^2 
  \left(\bk +\bp+ \bq \right)^2
 }
 \Bigg]\;.
 \label{eq:partial_fractions}
\end{align}
Now, we can write the entire $\mathbf{k}$-integrand as a sum of terms of the form%
\begin{align}
J_{\mu_1...\mu_M}\left(\mathbf{p}_1,\mathbf{p}_2,\mathbf{p}_3; n; {\nu_i}\right) =
\int d^n {\mathbf{k}} 
\frac{k_{\mu_1}...k_{\mu_M}}{
\left(\mathbf{k} + \mathbf{p}_1\right)^{2\nu_1}
\left(\mathbf{k} + \mathbf{p}_2\right)^{2\nu_2}
\left(\mathbf{k} + \mathbf{p}_3\right)^{2\nu_3}
}
\label{eq:pre_davy}
\end{align}
matching Eq.~(B.5) of Bzowski {\it et al.}\cite{bzowski12}. These tensor-valued integrals 
are now transformed into a permuted sum of scalar-valued integrals using Davydychev recursion 
relations \cite{davy91,davy92} as described in the Appendix of Ref.~\onlinecite{bzowski12}. 
Before the $\mathbf{q}$-integration, we obtain the intermediate result
\begin{align}
J (p)^{(4)} = \int_{\mathbf{q}}
\Bigg\{
\Bigg[&
   2 P^7 R+2 P^6 R (S-24 Q)-P^5
   \left(Q^2 (17 R+S)+40 Q R S+R^2 (5
   R+S)\right)
   \nonumber\\
   &
   -2 P^4 \left(76 Q^3 R+Q^2 S (4
   R+S)-12 Q R \left(2 R^2+S^2\right)+R^2 S
   (2 R+S)\right)+P^3 \Big(Q^4 (2 S-44 R)
   \nonumber\\
   &+2
   Q^3 S (S-29 R)+Q^2 \left(8 R^3+3 R^2 S+26
   R S^2-S^3\right)+2 Q R S \left(17 R^2+R
   S+10 S^2\right)
   \nonumber\\
   &
   +R^2 \left(4 R^3+R^2 S+2 R
   S^2-S^3\right)\Big)+P^2 \Big(-160 Q^5
   R+2 Q^4 S (S-30 R)
   \nonumber\\
   &
   +2 Q^3 \left(50 R^3+2
   R^2 S+23 R S^2+S^3\right)+Q^2 R S \left(3
   R^2+4 R S+13 S^2\right)
   \nonumber\\
   &
   -2 Q R^2 \left(6
   R^3-2 R^2 S+9 R S^2-S^3\right)+R^3 S
   \left(3 R^2+2 R S+S^2\right)\Big)-P
   \Big(Q^6 (21 R+S)
   \nonumber\\
   &
   +2 Q^5 S (7 R+S)+Q^4
   \left(-13 R^3-4 R^2 S-26 R
   S^2+S^3\right)-4 Q^3 R S \left(4 R^2+R S+3
   S^2\right)
   \nonumber\\
   &
   -Q^2 R^3 \left(R^2+5 R S-24
   S^2\right)+2 Q R^3 S \left(5 R^2-3 R S+6
   S^2\right)+R^4 \left(R^3+2 R
   S^2-S^3\right)\Big)
   \nonumber\\
   &
   +R \Big(-56 Q^7-46
   Q^6 S+2 Q^5 \left(26 R^2+R S+13
   S^2\right)+Q^4 S \left(31 R^2+4 R S+23
   S^2\right)
   \nonumber\\
   &
   -2 Q^3 R \left(6 R^3-R^2 S+8 R
   S^2-S^3\right)-2 Q^2 R^2 S \left(2 R^2-2 R
   S+5 S^2\right)+2 Q R^3 S^2 (S-R)
   \nonumber\\
   &
   -R^4 S
   \left(R^2+S^2\right)\Big)
\Bigg]
\times
\Bigg[
\frac{2}
   {N P
   Q^3 R S \left(P^2+Q^2-R^2\right)
   (P+Q+S)^2}
\Bigg]
\Bigg\}
\label{eq:pre_q}
\end{align}
To perform the second integration over $\mathbf{q}$, 
we re-write Eq.~(\ref{eq:pre_q}) as a two-dimensional integral over the angle variable $x=\frac{\mathbf{p}\cdot \mathbf{q}}{PQ}$ and 
the modulus $Q=|\mathbf{q}|$. These integrals diverge in the UV for large momenta $Q$. 
To separate the these UV-divergent terms, we expand the integrand around ``$Q=\infty$''. 

For the diverging terms, 
we introduce a UV cutoff $\Lambda$ and restrict the integration to values $Q\leq\Lambda$. There are two types of 
divergences: terms diverging as a power-law $\sim \Lambda$, where the integrand is a constant and terms 
diverging logarithmically, where the integrand is proportional to $P/Q$ (or $\sigma_0^2/Q$ in the symmetry-broken phase).
The linearly diverging terms are unimportant can be absorbed into constant counter-terms. 
In dimensional regularization, compliant with Lorentz symmetry, these would not be there anyway. On the other hand, 
the logarithmically divergent terms (e.g. those in Table \ref{tab:nu}) involve another energy scale and by the usual ``resummation'' of those terms, we can extract critical exponents.

The non-diverging terms can be integrated without a cutoff, sometimes even analytically but always numerically. Note that 
to compute the universal amplitudes $C_J$ and $C_A$ in Eqs.~(\ref{eq:value_cj_on},\ref{eq:value_cj_cp},\ref{eq:C_A}) 
\emph{all contributions} have to be carefully summed; it is not sufficient to restrict to singular terms.

We have programmed all of the 
just mentioned steps as a Mathematica algorithm to manage the computational complexity. 
Let us mention that the integration of {\emph{each}} of the multi-index terms of the form of Eq.~(\ref{eq:numerator}) 
can entail hundreds of terms which necessitates computerization of all the intermediate steps. 
Each individual substitution and permutation (sub-) routine was checked against direct numerical integration.

All other momentum integrals, including those for 3-loop diagrams (Fig.~\ref{fig:diags_gauge_crit}) or in the symmetry-broken 
phase (Fig.~\ref{fig:1loop}), can be reduced to products and sums of integrals of the form Eq.~(\ref{eq:pre_davy}) and we evaluate 
them similarly.

\end{document}